\documentclass[12pt,a4paper]{article}
\usepackage[text={16.5cm,23cm},centering]{geometry}
\usepackage{amsmath}
\usepackage{amsthm}
\usepackage{graphicx}
\graphicspath{{./}{fig/}}
\usepackage{tikz}
\usepackage[utf8]{inputenc}
\usepackage{tgtermes,tgheros,tgcursor}
\usepackage[varg]{newtxmath}
\usepackage{hyperref}
\hypersetup{%
 colorlinks=true,%
 linkcolor=blue,
 citecolor=blue,
}
\usepackage{xspace}
\newcommand{\floor}[1]{\left\lfloor #1\right\rfloor}
\numberwithin{equation}{section}
\newtheorem{lemma}{Lemma}
\newtheorem{theorem}[lemma]{Theorem}

\newcommand{\dd}{\mathrm{d}}
\newcommand{\Ncal}{\mathcal{N}}
\newcommand{\Zb}{\mathbb{Z}}
\newcommand{\Rb}{\mathbb{R}}
\newcommand{\Cb}{\mathbb{C}}

\DeclareMathOperator*{\tr}{\mathrm{tr}}

\newcommand{\Fs}{\mathsf{F}}
\newcommand{\Cs}{\mathsf{C}}
\newcommand{\Ps}{\mathsf{P}}
\newcommand{\csym}{\mathbb{Z}_{N}^{[1]}}
\newcommand{\Zg}{\widehat{Z}}
\newcommand{\U}{\mathrm{U}}
\newcommand{\Zbh}{\widehat{\mathbb{Z}}}
\DeclareMathOperator{\Hom}{\mathrm{Hom}}
\newcommand{\Tcal}{\mathcal{T}}
\newcommand{\Tch}{\widehat{\mathcal{T}}}
\newcommand{\wt}{\tilde{w}}
\DeclareMathOperator{\re}{\mathrm{Re}}
\newcommand{\ltl}{\tilde{\ell}}
\newcommand{\at}{\tilde{a}}
\newcommand{\ptl}{\tilde{p}}
\newcommand{\sba}{\bar{s}}
\newcommand{\lb}{\bar{l}}

\newcommand{\Ic}{\mathcal{I}}
\newcommand{\Jt}{\widetilde{J}}

\newcommand{\expval}[1]{\left\langle #1 \right\rangle}

\begin{document}
\begin{center}
  \begin{flushright}
    OU-HET 1333
  \end{flushright}
  \vspace{8ex}
  {\Large \bfseries \boldmath Non-invertible Duality Defects in Four-Dimensional $\mathbb{Z}_N$ Lattice Gauge Theories}\\
  \vspace{4ex}
  {\Large Satoshi Yamaguchi}\\
  \vspace{2ex}
  {\itshape Department of Physics, Graduate School of Science, 
  \\
  Osaka University, Toyonaka, Osaka 560-0043, Japan}\\
  \vspace{1ex}
  \begin{abstract}
    We consider four-dimensional $\mathbb{Z}_N$ lattice gauge theories with generalized plaquette actions. These theories are known to exhibit dualities. We construct topological interfaces implementing these dualities by assigning a weight to each building block and proving topological invariance. At self-dual couplings, these interfaces become duality defects representing non-invertible symmetries. We also study center symmetry defects and their junctions with the duality interfaces and derive several crossing relations.
  \end{abstract}
\end{center}

\tableofcontents

\section{Introduction}

Lattice field theories provide a conceptually straightforward and well-established framework for defining quantum field theories non-perturbatively.
In particular, lattice gauge theories \cite{Wilson:1974sk} preserve gauge symmetry exactly and have yielded many important numerical and theoretical results.
To take the continuum limit, one needs a second-order phase transition or a gapless phase. Thus, it is important to explore the phase structure of lattice gauge theories. For early work on these phase structures, see \cite{Balian:1974ts,Balian:1974ir,Balian:1974xw,Banks:1977cc,tHooft:1977nqb,Kogut:1979wt}.
In the study of the phase structure, it is important to consider symmetries.

$\Zb_N$ lattice gauge theories have played an important role in this development.
Early Monte Carlo studies by Creutz, Jacobs, and Rebbi clarified the phase structure of these gauge theories \cite{Creutz:1979kf,Creutz:1979zg}.
Analytical work also established that four-dimensional $\Zb_N$ models can support a massless photon phase for sufficiently large $N$ \cite{Elitzur:1979uv,Ukawa:1979yv,Frohlich:1982gf}.
More recently, the phase structure of $\Zb_N$ lattice gauge theories has been studied from a modern perspective \cite{Giansiracusa:2025hfj}.
These results make discrete gauge theories valuable models for understanding confinement and the emergence of continuous gauge fields at long distances.

Since the discovery of Kramers--Wannier (KW) duality \cite{Kramers:1941kn,Kramers:1941zz}, dualities have provided insight into phase transitions.
Wegner extended KW duality to general $\Zb_2$ models, including four-dimensional $\Zb_2$ lattice gauge theory \cite{Wegner:1971app}.
Further generalizations to four-dimensional $\Zb_N$ gauge theories were developed in Refs.~\cite{Casher:1978iz,Yoneya:1978dt,Fukugita:1982kk,Creutz:1982dn}. See also the review \cite{Savit:1979ny}.

More recently, the notion of symmetry has been extended by describing symmetries in terms of topological defects \cite{Gaiotto:2014kfa}, leading to the framework of ``generalized symmetries.''
In particular, generalized symmetries without a group structure are now called ``non-invertible symmetries.''
From this perspective, topological defects, or generalized symmetries, arising from KW duality in two dimensions at the self-dual point have been discovered and studied in many papers, including Refs.~\cite{Verlinde:1988sn,Frohlich:2004ef,Aasen:2016dop,Bhardwaj:2017xup,Chang:2018iay}.
Non-invertible topological defects based on KW-type dualities in four dimensions have been constructed \cite{Koide:2021zxj,Choi:2021kmx,Kaidi:2021xfk,Choi:2022zal}.
For reviews of recent developments in generalized symmetries, see, for example, Refs.~\cite{Shao:2023gho,Schafer-Nameki:2023jdn,Kaidi:2026urc}.

In this paper, we construct duality defects for $\Zb_N$ lattice gauge theories with generalized plaquette actions \cite{Drouffe:1978py,KorthalsAltes:1978tp,Fukugita:1982kk,Creutz:1982dn}.
We employ the Aasen--Mong--Fendley (AMF) approach \cite{Aasen:2016dop}, which has been used to construct duality defects in four-dimensional $\Zb_2$ lattice gauge theory in Ref.~\cite{Koide:2021zxj}.
In other words, we generalize the construction of \cite{Koide:2021zxj} to $\Zb_N$ lattice gauge theories with generalized plaquette actions.
We also study the half-space gauging approach \cite{Choi:2021kmx} and find that these two approaches give the same defects.

We also construct center symmetry defects and their junctions. We then derive several crossing relations involving these defects.

This paper is organized as follows.
In Sec.~\ref{sec:duality}, we set up $\Zb_N$ lattice gauge theories with generalized plaquette actions.
We then review the duality and generalize it to arbitrary nondegenerate symmetric bicharacters.
In Sec.~\ref{sec:duality-defects}, we construct duality defects. Here, we construct a duality interface between a pair of mutually dual theories.
In a self-dual theory, these duality interfaces become duality defects connecting identical theories.
We also derive universal identities that are independent of the details of the dynamics and are analogous to Ward--Takahashi identities.
In Sec.~\ref{sec:symmetry-junctions}, we construct center symmetry defects and their junctions. We derive crossing relations involving these defects and junctions.
In Sec.~\ref{sec:half-space-gauging}, we study the duality interfaces using half-space gauging. We find that the interfaces obtained here are the same as those obtained in Sec.~\ref{sec:duality-defects}.
Sec.~\ref{sec:conclusion} presents our conclusions and discussion.

\section{Dualities in four-dimensional \texorpdfstring{$\mathbb{Z}_N$}{ZN} lattice gauge theories}
\label{sec:duality}

In this section, we review and extend dualities of four-dimensional $\Zb_N$ lattice gauge theories with generalized actions \cite{Drouffe:1978py,KorthalsAltes:1978tp,Fukugita:1982kk,Creutz:1982dn}. We focus on the gauge group $\Zb_N$, but the construction can be generalized to an arbitrary finite abelian group.

\subsection{Finite Fourier transform}
We collect the notation and identities for the finite Fourier transform on $\Zb_N$.

For an integer $N>1$, we consider the group $\Zb_N$:
\begin{align}
  \Zb_N = \{0,1,2,\dots,N-1\}.
\end{align}
The group operation is addition, and all arithmetic is understood modulo $N$.

For $k\in\Zb$ with $\gcd(N,k)=1$, define the nondegenerate symmetric bicharacter $\chi_k: \Zb_N \times \Zb_N \to \mathrm{U}(1)$ by
\begin{align}
  \chi_k(a,b)=\exp\left(\frac{2\pi i k}{N} ab\right). \label{bichar}
\end{align}
The condition $\gcd(N,k)=1$ ensures nondegeneracy.
For $a,b,c\in\Zb_N$, we have
\begin{align}
  \chi_k(a,b+c)=\chi_k(a,b)\chi_k(a,c), \quad \chi_k(a,b)=\chi_k(b,a).\label{symbilin}
\end{align}
Intuitively, $\chi_k(a,b)$ is an exponentiated bilinear form.
For $b\in\Zb_N$, character orthogonality gives
\begin{align}
  \sum_{a\in \Zb_N}\chi_k(a,b) = N \delta_{b,0}. \label{chion}
\end{align}

A function $w:\Zb_N\to\Cb$ can be identified with a vector in $\Cb^N$.
Define the finite Fourier transform $\Fs_k$ by
\begin{align}
  (\Fs_k w)(a):=\frac{1}{\sqrt{N}}\sum_{b\in \Zb_N}\chi_k(a,b)w(b).
  \label{defFk}
\end{align}
Equivalently, $\Fs_k$ is the $N\times N$ matrix with entries
\begin{align}
  (\Fs_k)_{ab}=\frac{1}{\sqrt{N}}\chi_k(a,b).
\end{align}
By character orthogonality, $\Fs_k$ is unitary. Its square is
\begin{align}
  (\Fs_k^2 w)(a)=\frac{1}{N}\sum_{c\in\Zb_N}w(c)\sum_{b\in\Zb_N}\chi_k(b,a+c)
  =w(-a).
\end{align}
Here we used Eqs.~\eqref{symbilin} and~\eqref{chion}. Define the charge-conjugation operator $\Cs$ by
\begin{align}
  (\Cs w)(a):=w(-a).
\end{align}
Then $\Fs_k$ satisfies
\begin{align}
  \Fs_k^2=\Cs, \qquad \Cs^2=1,\qquad \Fs_k^4=1,\qquad
  \Fs_k^{-1}=\Fs_k^\dagger=\Fs_k^3.
\end{align}
For $N>2$, $\Fs_k$ has order four and generates a group isomorphic to $\Zb_4$. For $N=2$, $\Cs=1$ and $\Fs_k$ has order two.

We will use the following lemma later.
\begin{lemma}[Fourier transform preserves evenness]\label{lemma:Fourier}
Let $w:\Zb_N\to\Cb$ satisfy $w(-a)=w(a)$ for all $a\in\Zb_N$.
Then $\Fs_k w$ is also even:
\begin{align}
  (\Fs_k w)(-a)=(\Fs_k w)(a).
\end{align}
\end{lemma}
\begin{proof}
Changing the summation variable $b\mapsto-b$ gives
\begin{align}
  (\Fs_k w)(-a)
  &=\frac{1}{\sqrt{N}}\sum_{b\in\Zb_N}\chi_k(-a,b)w(b)\notag\\
  &=\frac{1}{\sqrt{N}}\sum_{b\in\Zb_N}\chi_k(a,b)w(-b)
  =(\Fs_k w)(a).
\end{align}
\end{proof}

\subsection{\texorpdfstring{$\Zb_N$}{ZN} lattice gauge theories}

We now define the four-dimensional $\Zb_N$ lattice gauge theories considered in this paper.  

Consider a finite four-dimensional hypercubic lattice $\Lambda$ with periodic boundary conditions. Choose a reference orientation for each link and plaquette.
Let $\Lambda_2$ denote the set of plaquettes, each taken once with its reference orientation.
We assign a variable $a_\ell\in\Zb_N$ to each oriented link $\ell$, with $a_{-\ell}=-a_\ell$ for the oppositely oriented link.
In the notation of Appendix~\ref{app:cochain}, $a\in C^1(\Lambda)$.
For a link $\ell=\expval{ij}$ oriented from site $i$ to site $j$, gauge transformations act as
\begin{align}
  a_\ell\longmapsto a_\ell+\lambda_j-\lambda_i,
  \label{eq:gauge-transformation}
\end{align}
where the gauge transformation parameter $\lambda_i\in\Zb_N$ is assigned independently to each site; equivalently, $\lambda\in C^0(\Lambda)$.
The gauge-invariant flux through an oriented plaquette $p$ is
\begin{align}
  f_p:=\sum_{\ell \in \partial p} a_{\ell},
  \label{eq:plaquette-flux}
\end{align}
where $\partial p$ consists of the four boundary links with the orientations induced by $p$. In the notation of Appendix~\ref{app:cochain}, $f=\dd a$.

Let $w:\Zb_N\to\Cb$ be a plaquette weight satisfying $\Cs w=w$, i.e.,
\begin{align}
  w(-f)=w(f),\qquad f\in\Zb_N.
  \label{eq:even-weight}
\end{align}
Since $f_{-p}=-f_p$, this condition makes the weight independent of the reference orientation of the plaquette.
The partition function is
\begin{align}
  Z_{w}
  :=\frac{1}{N^V}\sum_{a \in C^1(\Lambda)}\prod_{p\in\Lambda_2} w(f_p),
  \label{eq:partition-function}
\end{align}
where $V$ is the number of sites. The denominator $N^V$ represents the gauge volume.
Most of our analysis applies to general complex weights satisfying Eq.~\eqref{eq:even-weight}. For a probabilistic interpretation, the weights must be real and nonnegative.
For strictly positive weights, the action can be written as
\begin{align}
  S(a)=-\sum_{p\in\Lambda_2} \log w(f_p).
\end{align}
Then, the partition function is given by
\begin{align}
  Z_{w}=\frac{1}{N^V}\sum_{a \in C^1(\Lambda)}e^{-S(a)}.
\end{align}

\subsection{Gauging center symmetry}

The $\Zb_N$ lattice gauge theory considered in this paper has a one-form center symmetry $\csym$. We couple this symmetry to a background gauge field as follows.

The background gauge field for $\csym$ is given by $B\in C^2(\Lambda)$ with $\dd B=0$. A background gauge transformation is parametrized by $\lambda^1\in C^1(\Lambda)$, under which $a$ and $B$ transform as
\begin{align}
  a\longmapsto a + \lambda^1, \qquad B \longmapsto B+\dd \lambda^1.
\end{align}
Then, $f-B$ is gauge invariant, where $f=\dd a$.

The partition function in the presence of the background field $B$ is
\begin{align}
    Z_{w}(B)
  :=\frac{1}{N^V}\sum_{a \in C^1(\Lambda)}\prod_{p\in\Lambda_2} w(f_p-B_p). \label{PartitionFunction}
\end{align}

Gauging this symmetry amounts to summing over all flat configurations of the background gauge field. The partition function $\Zg_w$ of the gauged theory is
\begin{align}
  \Zg_w = \frac{1}{N^{3V}} \sum_{b \in C^2(\Lambda), \dd b = 0} Z_{w}(b).
\end{align}
Here, the denominator $N^{3V}$ is the gauge volume. The parameter of a one-form gauge transformation lies in $C^1(\Lambda)$, with a gauge-for-gauge redundancy parametrized by $C^0(\Lambda)$. Therefore, the gauge volume is given by
\begin{align}
  \frac{|C^1(\Lambda)|}{|C^0(\Lambda)|}=\frac{N^{4V}}{N^V}=N^{3V}.
\end{align}

This gauged theory has a one-form global symmetry, called the dual symmetry or quantum symmetry \cite{Gaiotto:2014kfa}. The dual one-form symmetry group is $\Zbh_N:=\Hom(\Zb_N, \U(1))$. Although $\Zbh_N\cong\Zb_N$, we must choose an identification between them. We use the isomorphism specified by the nondegenerate symmetric bicharacter $\chi_k$:
\begin{align*}
  \Zb_N\longrightarrow\Zbh_N,\qquad
  x\longmapsto \chi_k(x,\cdot).
\end{align*}

It is convenient to place the background gauge field for this dual one-form symmetry on the plaquettes of the dual lattice $\Lambda^*$.
Let $\Lambda_2^*$ denote the set of plaquettes of the dual lattice. There is a canonical bijection $\Lambda_2\longrightarrow\Lambda_2^*$, denoted by $p\mapsto p^*$.
Using the above identification, let $B\in C^2(\Lambda^*)$, $\dd B=0$, denote the background gauge field for the dual symmetry. The corresponding partition function is
\begin{align}
  \Zg_{w}(B)=\frac{1}{N^{3V}} \sum_{b \in C^2(\Lambda), \dd b = 0} Z_{w}(b) \prod_{p\in\Lambda_2}\chi_k(B_{p^*}, b_p). \label{gaugedPartitionFunction}
\end{align}

\subsection{Dualities}

We now state the duality for arbitrary $k$ coprime to $N$. For $k=1$, it reduces to the duality of Refs.~\cite{Fukugita:1982kk,Creutz:1982dn}.

Let $\Tcal_w$ denote the lattice gauge theory with weight $w$, and let $\Tch_w$ denote the theory obtained by gauging the center symmetry $\csym$ of $\Tcal_w$.
Let $\wt$ be the Fourier transform of $w$:
\begin{align}
  \wt := \Fs_k w.  \label{wdual}
\end{align}
By Lemma~\ref{lemma:Fourier}, $\wt$ is also even, so it defines a lattice gauge theory $\Tcal_{\wt}$.
The following equivalence holds:
\begin{align}
  \Tcal_{\wt}\cong\Tch_{w}.
\end{align}
At the level of partition functions, this equivalence takes the following form.
\begin{theorem}[Duality]
  If $w$ and $\wt$ are related by Eq.~\eqref{wdual}, the partition functions defined in Eqs.~\eqref{PartitionFunction} and~\eqref{gaugedPartitionFunction} satisfy
  \begin{align}
  Z_{\wt}(B)=\Zg_{w}(B).\label{dualityPartitionFunction}
  \end{align}
\end{theorem}
\begin{proof}
  In Eq.~\eqref{gaugedPartitionFunction}, we replace the sum over flat $b$ by a sum over all $b\in C^2(\Lambda)$, with an insertion of $\delta_{\dd b,0}$. The flatness constraint factorizes as
  \begin{align}
    \delta_{\dd b,0}=\prod_{c\in \Lambda_3}\delta_{(\dd b)_c,0}.
  \end{align}
  Using Eq.~\eqref{chion}, we can rewrite $\delta_{(\dd b)_c,0}$ as
  \begin{align}
    \delta_{(\dd b)_c,0} = \frac{1}{N} \sum_{g_{c^{*}}\in \Zb_N}\chi_k(g_{c^*},(\dd b)_c),
  \end{align}
  where we use the bijection $\Lambda_3 \to \Lambda_1^*$ given by $c\mapsto c^*$ and introduce a dual one-cochain $g \in C^1(\Lambda^*)$.  Then, we can rewrite $\delta_{\dd b,0}$ as
  \begin{align}
    \delta_{\dd b,0}=\sum_{g\in C^1(\Lambda^*)}\prod_{c\in \Lambda_3}\frac{1}{N}\chi_k(g_{c^*},(\dd b)_c)
    =\frac{1}{N^{4V}}\sum_{g\in C^1(\Lambda^*)} \prod_{p \in \Lambda_2}\chi_k(-(\dd g)_{p^{*}},b_p).
    \label{deltadb}
  \end{align}
  Here, we use $|\Lambda_3|=4V$ and Lemma \ref{lemma:cochain-integration-by-parts}.
  To eliminate the sum over $a$, first substitute Eq.~\eqref{PartitionFunction} into Eq.~\eqref{gaugedPartitionFunction}:
  \begin{align*}
    \Zg_w(B)=\frac{1}{N^{4V}}\sum_{a\in C^1(\Lambda)}
    \sum_{\substack{b\in C^2(\Lambda)\\ \dd b=0}}
    \prod_{p\in\Lambda_2}w((\dd a)_p-b_p)\chi_k(B_{p^*},b_p).
  \end{align*}
  For each $a$, make the change of variables $b'=b-\dd a$, which preserves the flatness constraint. This is equivalent to setting $a=0$ by the one-form gauge transformation with parameter $\lambda^1=-a$.
  Since $\dd B=0$, Lemma~\ref{lemma:cochain-integration-by-parts} gives
  \begin{align*}
    \prod_{p\in\Lambda_2}\chi_k(B_{p^*},(\dd a)_p)=1.
  \end{align*}
  Since $w(-b'_p)=w(b'_p)$, the summand is independent of $a$. The sum over $a$ therefore contributes $|C^1(\Lambda)|=N^{4V}$, canceling the prefactor. Relabeling $b'$ as $b$, we obtain
  \begin{align*}
    \Zg_w(B)=\sum_{\substack{b\in C^2(\Lambda)\\ \dd b=0}}
    \prod_{p\in\Lambda_2}w(b_p)\chi_k(B_{p^*},b_p).
  \end{align*}
  Inserting Eq.~\eqref{deltadb} then gives
  \begin{align}
    \Zg_{w}(B)&
    =\frac{1}{N^{4V}}\sum_{g\in C^1(\Lambda^*)}
    \sum_{b \in C^2(\Lambda)} \prod_{p \in \Lambda_2} w(b_p) \chi_k(B_{p^*},b_p)\chi_k(-(\dd g)_{p^{*}},b_p)\notag \\
    &=\frac{1}{N^{4V}}\sum_{g\in C^1(\Lambda^*)}
    \prod_{p \in \Lambda_2} \sum_{b_p \in \Zb_N}  w(b_p) \chi_k(B_{p^*}-(\dd g)_{p^*},b_p)\notag\\
    &=\frac{1}{N^{4V}}\sum_{g\in C^1(\Lambda^*)}
    \prod_{p \in \Lambda_2} \sqrt{N}\wt(B_{p^*}-(\dd g)_{p^*})\notag\\
    &=\frac{1}{N^{V}}\sum_{g\in C^1(\Lambda^*)}
    \prod_{p^* \in \Lambda^*_2}\wt((\dd g)_{p^*}-B_{p^*}).
    \label{dualityfinal}
  \end{align}
  Here, we use Eq.~\eqref{symbilin}, $\wt=\Fs_k w$ with Eq.~\eqref{defFk}, $|\Lambda_2|=6V$, and $\wt(a)=\wt(-a)$.  The final form of Eq.~\eqref{dualityfinal} is identical to \eqref{PartitionFunction} with $w$ replaced by $\wt$ and $\Lambda$ replaced by $\Lambda^{*}$.  Since the hypercubic lattices $\Lambda$ and $\Lambda^*$ are isomorphic to each other, Eq.~\eqref{dualityfinal} implies Eq.~\eqref{dualityPartitionFunction}.
\end{proof}

This duality exchanges the original lattice $\Lambda$ and its dual $\Lambda^*$. This exchange guides our construction of duality defects in Sec.~\ref{sec:duality-defects}.

\subsection{Space of self-dual theories}
\label{sec:self-dual-theories}

In this subsection, we study exactly self-dual weights, defined by $\Fs_k w=w$, and count their independent parameters modulo overall rescaling. We first compute the complex dimension of the space of exactly self-dual weights. We then show that its strictly positive real subset has the same dimension over $\Rb$.
For a nonzero even weight, self-duality up to an overall rescaling, $\Fs_k w=cw$, allows $c=\pm1$, since $\Fs_k^2=\Cs$. We restrict the complex and real weights below to the $+1$ eigenspace. For strictly positive weights, $(\Fs_k w)(0)=N^{-1/2}\sum_a w(a)>0$ and $w(0)>0$ force $c=1$, so this restriction does not exclude any self-dual positive weights.

We begin with the space $\mathrm{Map}(\Zb_N, \Cb) \cong \Cb^N$. The exactly self-dual weights form the eigenspace $V_+$ of $\Fs_k$ with eigenvalue $1$. Notice that the condition \eqref{eq:even-weight} is automatically satisfied by every weight in this space. In order to compute the dimension of $V_+$, it is convenient to introduce the projector onto $V_+$:
\begin{align}
  \Ps:=\frac{1}{4}(1+\Fs_k+\Fs_k^2+\Fs_k^3)
  =\frac{1}{4}(1+\Cs+\Fs_k+\Fs_k^{\dag}).
  \label{self-dual-projection}
\end{align}
Then, $\dim_{\Cb} V_+ = \tr \Ps$.

Clearly, $\tr 1=N$. Since $(\Cs w)(a)=w(-a)$, $\tr \Cs$ equals the number of solutions to $a=-a$ in $\Zb_N$. Therefore,
\begin{align}
  \tr \Fs_k^2 =\gcd(N,2)=
  \begin{cases}
    1 & (N:\text{ odd})\\
    2 & (N:\text{ even})
  \end{cases}.
\end{align}

The remaining part is evaluated as follows.
\begin{align}
  \tr(\Fs_k+\Fs_k^{\dag})=2 \re \tr \Fs_k = \frac{2}{\sqrt{N}}\re G(k;N),\\
  G(k;N):=\sum_{r=1}^{N}\exp\left(\frac{2\pi i k r^2}{N}\right).
\end{align}
The quantity $G(k;N)$ is a quadratic Gauss sum.

Combining these results, we obtain
\begin{align}
  \dim_{\Cb} V_+ = \frac14\left(
    N+\gcd(N,2)+\frac{2}{\sqrt{N}}\re G(k;N)
  \right).
\end{align}
For $c\in\Cb^\times:=\Cb \setminus \{0\}$, rescaling the weight changes the partition function by
\begin{align*}
  Z_{cw}=c^{|\Lambda_2|}Z_w=c^{6V}Z_w,
\end{align*}
where $V$ is the number of sites. We disregard this overall normalization and identify nonzero weights in $V_+$ that differ by an overall rescaling. The resulting parameter space is $\mathbb{P}(V_+)=(V_+\setminus\{0\})/\Cb^\times$, with complex dimension $d_{N,k}=\dim_{\Cb}V_+-1$. Thus,
\begin{align}
  d_{N,k}=\frac12 \floor{\frac{N}{2}}+\frac{1}{2\sqrt{N}}\re G(k;N)-\frac12.
\end{align}  
To evaluate this number, we need to evaluate the Gauss sum $G(k;N)$. The expression for $d_{N,k}$ in terms of the Jacobi symbol is given by Eq.~\eqref{eq:dNk-general-Jacobi}. The derivation of this result from the formulas in Ref.~\cite{Apostol} is given in Appendix~\ref{app:GaussSum}.
When $k=1$, $d_{N,1}=\floor{\frac{N}{4}}$, in agreement with Refs.~\cite{Fukugita:1982kk,Creutz:1982dn}.  In general, $d_{N,k}=\floor{\frac{N}{4}}$ or $d_{N,k}=\floor{\frac{N}{4}}-1$ depending on $N,k$.

We now consider the space of self-dual theories with positive weights.
Since $\Fs_k$ is a symmetric unitary matrix and $\Fs_k^3=\Fs_k^{\dag}$, the projector $\Ps$ in Eq.~\eqref{self-dual-projection} is real.
Therefore, the space of exactly self-dual real weights is $V_{+,\Rb}:=V_+\cap\Rb^N=\Ps \Rb^N$. Its real dimension is $\dim_{\Rb}V_{+,\Rb}=d_{N,k}+1$.
Identifying nonzero weights that differ by a nonzero real rescaling gives a family of real dimension $d_{N,k}$. To determine the dimension for strictly positive weights, consider the following self-dual weight:
\begin{align}
  w_{P}(0)=1+\sqrt{N}, \qquad w_{P}(a)=1,\quad a\ne 0.
\end{align}
Its self-duality follows from the representation
\begin{align}
  w_P=w_1 + \Fs_k w_1,\qquad w_1(a):=1 , \ \forall a \in \Zb_N,\qquad \Fs_k^2 w_1=w_1.
\end{align}
Strict positivity is an open condition in $V_{+,\Rb}$. Since $w_P(a)>0$ for every $a$, a sufficiently small neighborhood of $w_P$ in $V_{+,\Rb}$ consists entirely of strictly positive weights. Thus, the cone of strictly positive self-dual weights has real dimension $d_{N,k}+1$. Quotienting by positive rescalings reduces the dimension by one, giving $d_{N,k}$.

\subsection{Examples of duality and self-duality}
\label{sec:duality-examples}

We now give some examples of the Fourier duality in Eq.~\eqref{wdual} and the self-dual parameter spaces of Sec.~\ref{sec:self-dual-theories} with positive plaquette weights.
We use $k=1$ unless otherwise stated.
Since $\Fs_{-1}w=\Fs_1w$ for even weights, the results for $N=2,3,4$ apply to every allowed $k$.

\paragraph{Wilson weights for $N=2,3$.}
Consider the Wilson plaquette weight \cite{Wilson:1974sk}
\begin{align}
  w_K(a)=\exp\left[K\cos\left(\frac{2\pi a}{N}\right)\right],
  \qquad K>0.
  \label{eq:example-wilson-weight}
\end{align}
For $N=2$, its Fourier transform is
\begin{align}
  \Fs_1w_K&=\sqrt{\sinh(2K)}\,w_{\widetilde K},
  & e^{-2\widetilde K}&=\tanh K,
  & \sinh(2K)\sinh(2\widetilde K)&=1.
  \label{eq:example-z2-duality}
\end{align}
This is the four-dimensional Kramers--Wannier--Wegner duality \cite{Wegner:1971app}.
The positive fixed point is
\begin{align}
  K_{\mathrm{sd}}^{(2)}=\frac12\log(1+\sqrt2).
\end{align}
At this value the prefactor in Eq.~\eqref{eq:example-z2-duality} is one, so the weight is exactly self-dual.

For $N=3$, write $r=e^{3K/2}$, so that
$w_K=e^{-K/2}(r,1,1)$, with entries ordered by $a=0,1,2$.
Direct Fourier transformation gives
\begin{align}
  \Fs_1w_K
  =\frac{e^{-K/2}}{\sqrt3}(r+2,r-1,r-1).
  \label{eq:example-z3-fourier}
\end{align}
Up to an overall factor, the transformed weight again has the Wilson form, with
\begin{align}
  e^{3\widetilde K/2}=\frac{r+2}{r-1},
  \qquad
  (e^{3K/2}-1)(e^{3\widetilde K/2}-1)=3.
  \label{eq:example-z3-duality}
\end{align}
The positive fixed point is therefore
\begin{align}
  K_{\mathrm{sd}}^{(3)}=\frac23\log(1+\sqrt3).
\end{align}
Substituting $r=1+\sqrt3$ in Eq.~\eqref{eq:example-z3-fourier} verifies exact self-duality, including the normalization.
These $N=2,3$ self-dual weights are proportional to the weight $w_P$ introduced above; they are the unique strictly positive self-dual weights up to rescaling, consistent with $d_{2,1}=d_{3,1}=0$.

\paragraph{The one-parameter self-dual family for $N=4$.}
For an arbitrary even weight $w=(u,x,t,x)$, the Fourier transform reads
\begin{align}
  \Fs_1w=\frac12(u+2x+t,\,u-t,\,u-2x+t,\,u-t).
  \label{eq:example-z4-fourier}
\end{align}
Consequently, exact self-duality is equivalent to $u=t+2x$.
With the normalization $u=1$, all strictly positive solutions have the form
\begin{align}
  w=(1,x,1-2x,x),\qquad 0<x<\frac12.
  \label{eq:example-z4-family}
\end{align}
This realizes the parameter count $d_{4,1}=1$ explicitly.
In the generalized action studied in Refs.~\cite{Fukugita:1982kk,Creutz:1982dn}, we may parametrize the weight, up to rescaling, by
\begin{align}
  w_{K_1,K_2}(a)
  =\exp\left[K_1\cos\left(\frac{\pi a}{2}\right)+K_2\cos(\pi a)\right].
\end{align}
Then $(u,x,t)=(e^{K_1+K_2},e^{-K_2},e^{-K_1+K_2})$, and the self-dual family is the curve
\begin{align}
  e^{2K_2}\sinh K_1=1,\qquad K_1>0.
  \label{eq:example-z4-self-dual-curve}
\end{align}
It contains the Wilson point
$(K_1,K_2)=(\log(1+\sqrt2),0)$ and the Potts-type point
$(K_1,K_2)=(\tfrac12\log3,\tfrac14\log3)$, where $w$ is proportional to $(3,1,1,1)$.

The Wilson family is closed under duality up to an overall rescaling.
Setting $K_2=0$ and $K_1=K>0$ in Eq.~\eqref{eq:example-z4-fourier} gives
\begin{align}
  \Fs_1w_K
  &=(\cosh K+1,\sinh K,\cosh K-1,\sinh K)
    =\sinh K\,w_{\widetilde K},\notag\\
  \widetilde K&=\log\coth(K/2),
  \qquad \sinh K\sinh\widetilde K=1.
  \label{eq:example-z4-wilson-duality}
\end{align}
Both $w_K$ and $\Fs_1w_K$ are strictly positive for every $K>0$.
The two weights coincide precisely at $K=\log(1+\sqrt2)$, where the prefactor $\sinh K$ is one.

\paragraph{Villain weights for arbitrary $N$.}
A family closed under the standard Fourier duality up to rescaling for every $N$ is provided by the Villain periodic Gaussian \cite{Villain:1974ir}, applied to four-dimensional $\Zb_N$ gauge theory by Casher \cite{Casher:1978iz}.
It is convenient to use a parameter $\tau_2>0$ and define\footnote{The weights are defined by a theta function with modular parameter $\tau=i\tau_2$.}
\begin{align}
  v_{\tau_2}(a):=\sum_{m\in\Zb}
    \exp\left[-\frac{\pi\tau_2}{N}(a+Nm)^2\right].
  \label{eq:example-villain-weight}
\end{align}
This weight is even and strictly positive.
Combining the sum over $a\in\Zb_N$ with the integer sum and applying the Poisson resummation formula gives, for any $k$ coprime to $N$,
\begin{align}
  (\Fs_kv_{\tau_2})(b)
  &=\frac{1}{\sqrt N}\sum_{a \in \Zb_N}\sum_{m\in\Zb}
    \exp\left[-\frac{\pi\tau_2}{N}(a+Nm)^2+\frac{2\pi i k b a}{N}\right]\notag\\
  &=\frac{1}{\sqrt N}\sum_{j\in\Zb}
    \exp\left[-\frac{\pi\tau_2 j^2}{N}+\frac{2\pi i k b j}{N}\right]\notag\\
  &=\frac{1}{\sqrt{\tau_2}}\sum_{n\in\Zb}
    \exp\left[-\frac{\pi(kb-Nn)^2}{N\tau_2}\right]
    =\tau_2^{-1/2}v_{1/\tau_2}(kb).
  \label{eq:example-villain-duality}
\end{align}
For $k=1$, duality therefore sends $\tau_2$ to $1/\tau_2$, and $v_1$ is exactly self-dual.
The duality defect of this self-dual lattice model is also discussed in Ref.~\cite{Choi:2021kmx}.
For general $k$, Eq.~\eqref{eq:example-villain-duality} still gives a positive dual pair, but self-duality at $\tau_2=1$ additionally requires $v_1(kb)=v_1(b)$; it holds for $k=\pm1$ and cannot be assumed for an arbitrary bicharacter.

\section{Duality defects}
\label{sec:duality-defects}

In this section, we construct duality interfaces by extending the lattice construction of Ref.~\cite{Koide:2021zxj} to $\Zb_N$ gauge theories.
We consider two even plaquette weights related by
\begin{align}
  \wt=\Fs_k w.
  \label{eq:interface-dual-weights}
\end{align}
The interface separates regions described by $w$ and $\wt$, with the roles of the original and dual lattices exchanged across it.
When $w=\wt$, it becomes a duality defect within a single theory.
The positive dual pairs and self-dual weights in Sec.~\ref{sec:duality-examples} provide concrete examples of both cases.
We first describe its geometry, then specify the local weights and establish gauge invariance and invariance under local deformations.
Finally, we study the expectation value of a spherical defect and the effect of taking connected sums.

\subsection{Geometry}
\label{sec:defect-geometry}

We use the lattice geometry of Ref.~\cite{Koide:2021zxj}.
Choose the lattice spacing to be two and realize the sites of the original and dual lattices on the covering space $\Rb^4$ as
\begin{align}
  \Lambda=(2\Zb)^4,\qquad
  \Lambda^*=(2\Zb+1)^4.
  \label{eq:two-lattice-sites}
\end{align}
For a finite periodic lattice, we identify coordinates with compatible even periods.
Within each lattice, nearest-neighbor sites are joined by links, and elementary squares are plaquettes.
The lattice carrying gauge variables in a given bulk region is called \emph{active}; the other is called \emph{inactive}.

\begin{figure}
  \centering
  \includegraphics[width=7cm]{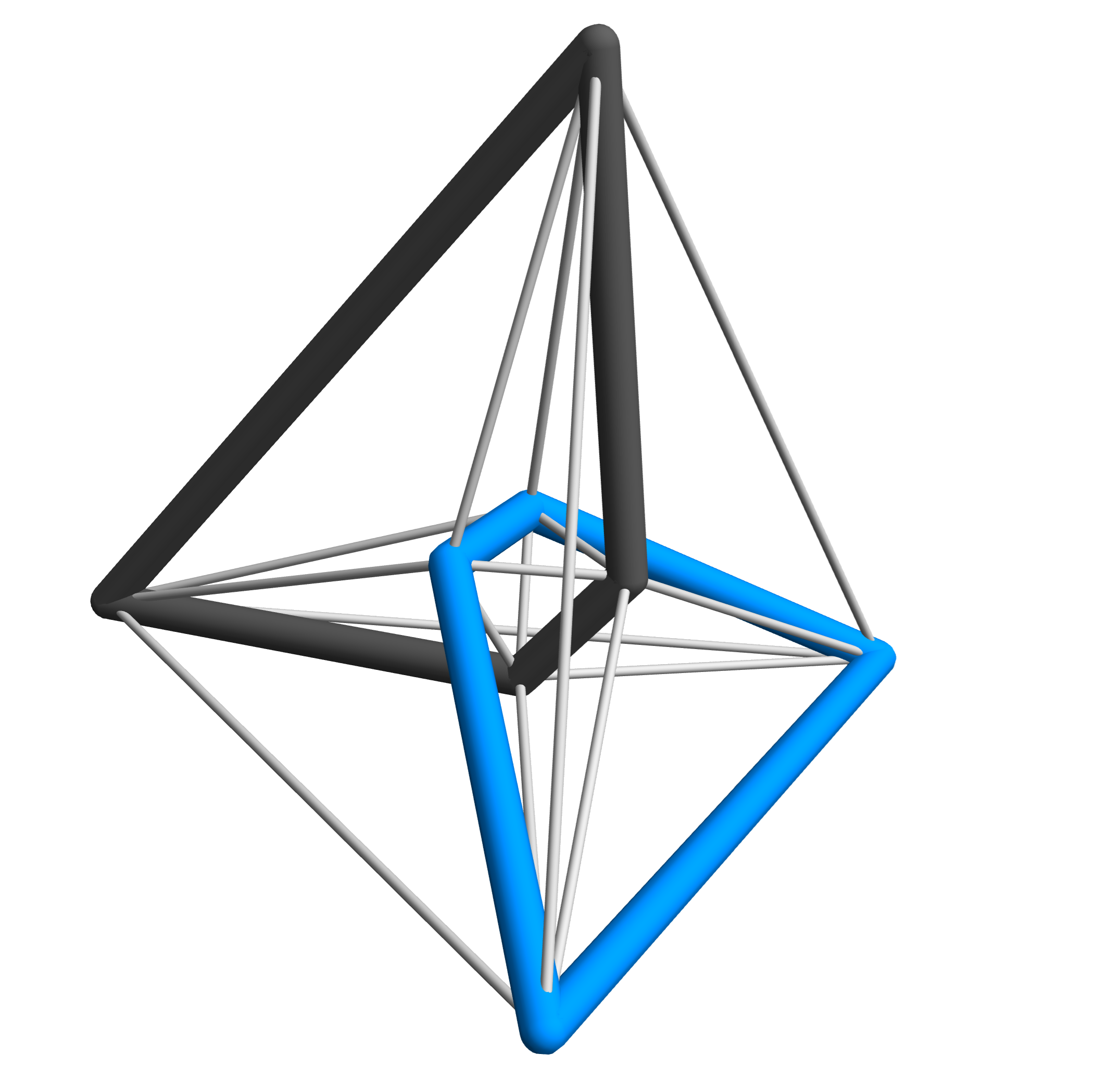}
  \caption{The 16-cell is the convex hull of the two plaquettes shown. The black plaquette is active and the blue one is inactive.}\label{fig:16-cell}
\end{figure}

For each plaquette $p\in\Lambda_2$, its dual plaquette $p^*\in\Lambda_2^*$ has the same center and lies in the orthogonal two-plane.
The convex hull of $p$ and $p^{*}$ is a regular 16-cell, a four-dimensional polytope with eight vertices and sixteen tetrahedral facets (see Fig.~\ref{fig:16-cell}).
For example, consider two plaquettes with the following vertices:
\begin{align}
  p&=(0,0,0,0),(2,0,0,0),(2,2,0,0),(0,2,0,0),\notag\\
  p^*&=(1,1,-1,-1),(1,1,1,-1),(1,1,1,1),(1,1,-1,1),
  \label{eq:plaquette-pair-example}
\end{align}
whose common center is $(1,1,0,0)$.
These 16-cells tessellate four-dimensional space, with one 16-cell for each pair $p,p^*$.
In particular, the plaquette weight can be regarded as a weight assigned to the corresponding 16-cell.

The facets of the 16-cell associated with the pair $p,p^*$ are described as follows.  Let $\ell_i$, $i=1,\ldots,4$, be the links around $p$, and let $\ltl_j$, $j=1,\ldots,4$, be those around $p^*$.
Each tetrahedral facet contains one link of each lattice as a pair of opposite edges. We denote the tetrahedron containing $\ell_i$ and $\ltl_j$ by $(i,j)$.



We divide four-dimensional space into two regions, the $w$ region and the $\wt$ region, separated by a closed surface $M$ made of the tetrahedral facets of 16-cells.
Each boundary element is duplicated, with one copy placed on each side.  In other words, the boundary is made of tetrahedral prisms.
We take $\Lambda$ to be active and $\Lambda^*$ to be inactive in the $w$ region, while these assignments are reversed in the $\wt$ region.
Consequently, a prism carries two active link variables, one from each side, as shown in Fig.~\ref{fig:tetrahedralprism}.


\subsection{Weights for the duality defects}

\begin{figure}
  \centering
  \includegraphics[width=8cm]{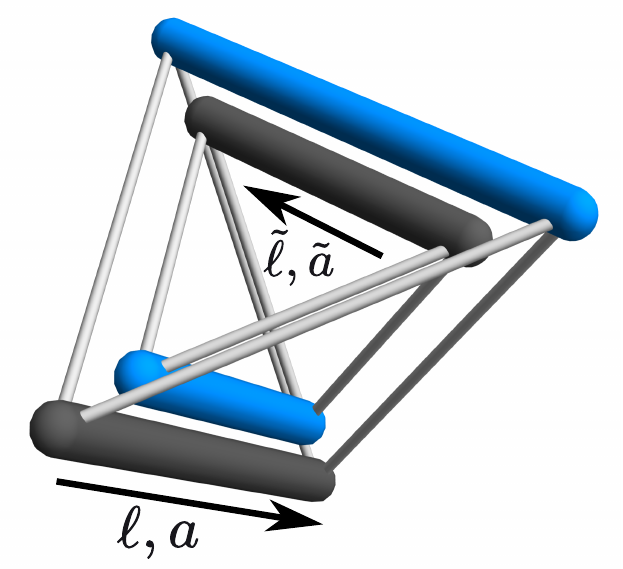}
  \caption{A tetrahedral prism supporting a duality interface. The outer tetrahedron belongs to the $w$ region, while the inner one belongs to the $\wt$ region. The thick black links are active links to which degrees of freedom are assigned. The arrows indicate the standard orientations, which determine the sign convention for the link variables in the interface weight.}
  \label{fig:tetrahedralprism}
\end{figure}

We describe duality interfaces by assigning an appropriate weight to each tetrahedral prism at the boundary between the two regions, in addition to the bulk Boltzmann weights.  We also assign constant weights to active and inactive sites and links.

The Boltzmann weights are assigned to each active plaquette, or equivalently to each 16-cell: $w(f_p)$ for a plaquette $p$ in the $w$ region and $\wt(f_{\ptl})$ for a plaquette $\ptl$ in the $\wt$ region.

The weight for a tetrahedral prism is denoted by $D(a,\at)$, with the link variables $a,\at$ oriented as in Fig.~\ref{fig:tetrahedralprism}.
The weights for an active site, an active link, an inactive site, and an inactive link are denoted by $s,l,\sba, \lb$, respectively.
In this paper, we consider the following weights:
\begin{align}
   D(a,\at)=\chi_{k}(a,\at), \quad
   s=l=\frac{1}{\sqrt{N}},\quad \sba=\lb=1.
   \label{dualitydefectweight}
\end{align}

First, let us check the gauge invariance of the interface.  Although the weight $D(a,\at)$ for a building block is not gauge invariant, the total weight for the interface on a closed surface is gauge invariant.  As shown in Fig.~\ref{fig:tetrahedralprism}, for an active site $i$ in a tetrahedron, there is a triangle $t$ that contains the active site $i$ and an inactive link.  The triangle $t$, together with the site $i$, is shared by two tetrahedral prisms in the closed surface. In these two tetrahedra, the two active links connected to $i$ are oppositely oriented.  The factors induced by a gauge transformation at $i$ therefore cancel between these two links.

\subsection{Proof of topological invariance}
\label{sec:defect-topological-invariance}

Here, we prove that the interface described by the weights \eqref{dualitydefectweight} is topological.

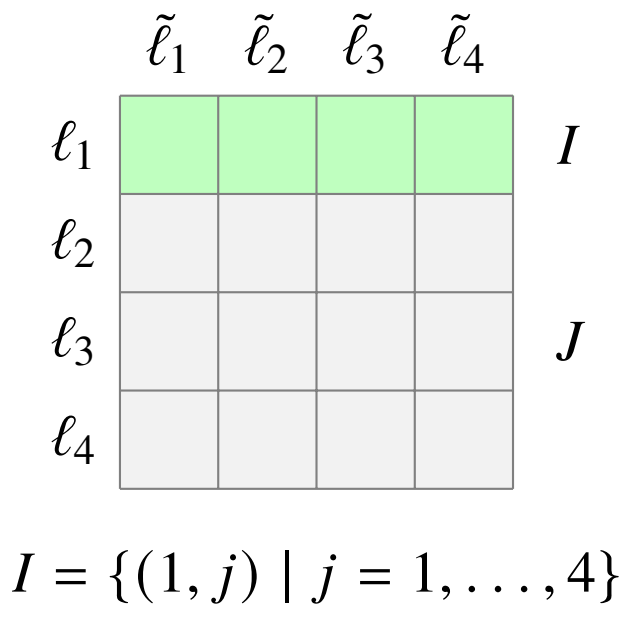
\begin{figure}[htbp]
  \centering
  \begin{tikzpicture}[font=\small]
    \begin{scope}[xshift=1.8cm,yshift=1.3cm,scale=0.65]
      \fill[gray!10] (0,0) rectangle (4,-4);
      \fill[green!25] (0,0) rectangle (4,-1);
      \draw[step=1,gray] (0,0) grid (4,-4);
      \foreach \i in {1,...,4}{
        \node[left] at (0,-\i+0.5) {$\ell_\i$};
        \node[above] at (\i-0.5,0) {$\ltl_\i$};
      }
      \node[right] at (4.2,-0.5) {$I$};
      \node[right] at (4.2,-2.5) {$J$};
      \node[below] at (2,-4.4) {$I=\{(1,j)\mid j=1,\ldots,4\}$};
    \end{scope}
  \end{tikzpicture}
  \caption{An array indexing the sixteen facets, with an example of complementary patches $I$ and $J$.
  Each entry represents a tetrahedron, rather than a square in space; both displayed patches are three-balls.}
  \label{fig:defect-geometry}
\end{figure}

\begin{figure}
  \centering
  \includegraphics[width=8cm]{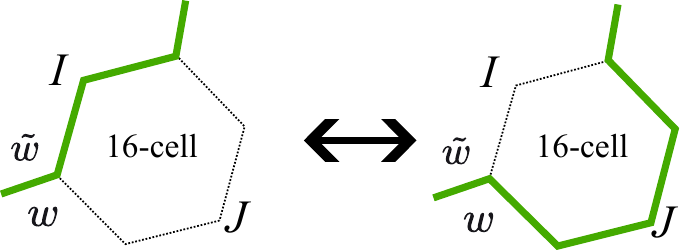}
  \caption{Schematic picture of a local deformation.}
  \label{fig:localdeformation}
\end{figure}

Let us explain a local deformation across a 16-cell (see Fig.~\ref{fig:localdeformation}).
Label its tetrahedral facets by
\begin{align}
  U:=\{(i,j)\mid i,j=1,\ldots,4\}.
\end{align}
$(i,j)$ denotes the tetrahedron that contains $\ell_i$ and $\ltl_j$.
We require the closed 16-cell $P$ to be an embedded four-ball in the periodic lattice.
For a patch $I\subset U$ and its complement $J:=U\setminus I$, let $|I|$ and $|J|$ denote the respective unions of closed tetrahedra.
We allow the move when $M\cap P=|I|$ as sets and $|I|$ and $|J|$ are nonempty piecewise-linear three-balls meeting along their common boundary two-sphere.
Replacing $|I|$ by $|J|$ then attaches a three-ball to the unchanged part of $M$ along the same boundary and preserves the topology of $M$.
An example is shown in Fig.~\ref{fig:defect-geometry}.
The two patches carry opposite signs relative to the orientation induced from the 16-cell, so that their orientations match the unchanged part of $M$.

As the interface passes through the 16-cell, its active and inactive plaquettes are exchanged, together with the bulk weight $w \leftrightarrow \wt$.
Topological invariance requires equality of the local contributions to the partition function before and after this move, with all external active link variables held fixed and any internal active link variables summed over.
These equalities are the \emph{defect commutation relations}; we will verify them for the weights and normalization in Eq.~\eqref{dualitydefectweight}.
Covering the entire 16-cell creates a closed spherical component and is not a topology-preserving move of this type.

\paragraph{Fourier identities and orientation conventions.}
The algebraic input is the pair of identities
\begin{align}
  \sum_{u\in\Zb_N}w(u+X)\chi_k(u,Y)
  &=\sqrt{N}\,\wt(Y)\chi_k(-X,Y),
  \label{eq:defect-forward-fourier}\\
  \sum_{v\in\Zb_N}\wt(v+Y)\chi_k(-X,v)
  &=\sqrt{N}\,w(X)\chi_k(X,Y).
  \label{eq:defect-reverse-fourier}
\end{align}
The first follows by using $u+X$ as the summation variable and applying Eq.~\eqref{defFk}.
The second follows in the same way from $w=\Fs_{-k}\wt$, since character orthogonality gives $\Fs_{-k}=\Fs_k^{-1}$.

Let $p$ and $\ptl$ denote the cycles of two plaquettes in the 16-cell:
\begin{align}
  p=\expval{\ell_1 \ell_2 \ell_3 \ell_4},\quad
  \ptl = \expval{\ltl_1 \ltl_2 \ltl_3 \ltl_4}.
  \label{eq:defect-boundary-join}
\end{align}
Orient the links along these cycles and denote their variables by $a_i$ and $\at_j$, respectively.
Choose the cycle orientations so that the tetrahedra in the initial patch $I$ have the standard orientation shown in Fig.~\ref{fig:tetrahedralprism}.
The complementary patch $J$ then has the opposite sign.
Reversing a cycle orientation leaves its bulk weight unchanged by evenness.

For later use, let us define
\begin{align}
  X:=\sum_{i=1}^4 a_i,\qquad Y:=\sum_{j=1}^4\at_j,\qquad
  H_I:=\prod_{(i,j)\in I}\chi_k(a_i,\at_j),\qquad
  H_J:=\prod_{(i,j)\in J}\chi_k(a_i,\at_j).
  \label{eq:defect-patch-factors}
\end{align}
By the property \eqref{symbilin}, the following identity holds:
\begin{align}
  H_I H_J=\chi_k(X,Y).
  \label{eq:defect-complementary-factors}
\end{align}
Apart from the constant site and link weights, the local contributions before and after the move are therefore $w(X)H_I$ and $\wt(Y)H_J^{-1}$.
The inverse on $H_J$ is essential: it accounts for the opposite orientation of the complementary patch.

\paragraph{Internal links and the geometry of an allowed move.}
Let
\begin{align}
  R&:=\{i\mid (i,j)\in I\text{ for every }j\},\qquad r:=|R|,\notag\\
  C&:=\{j\mid (i,j)\in J\text{ for every }i\},\qquad c:=|C|.
  \label{eq:defect-internal-links}
\end{align}
Thus, $R$ consists of the complete rows of $I$, and $C$ consists of its empty columns.  For example, for the configuration of Fig.~\ref{fig:defect-geometry}, $R=\{1\}$ and $C=\emptyset$.
The internal active links before the move are precisely the $a_i$ with $i\in R$: all four facets incident on such a link lie in $I$, and the link disappears from the active lattice after the move.
Similarly, the internal active links after the move are the $\at_j$ with $j\in C$.
All other active link variables are external and are held fixed.
A complete row and an empty column cannot coexist: $r>0$ implies $c=0$, while $c>0$ implies $r=0$.

\begin{lemma}\label{lemma:defect-ball-move}
For every allowed move, either $r>0$ or $c>0$.
\end{lemma}
\begin{proof}
The two cycles $p$ and $\ptl$ in \eqref{eq:defect-boundary-join} form a Hopf link on the surface of the 16-cell, as shown in Fig.~\ref{fig:16-cell}.
The linking number of the Hopf link has absolute value one.
Suppose $r=c=0$.
Since every column meets $I$, the entire cycle $\ptl$ lies in $I$.
Since every row meets $J$, the entire cycle $p$ lies in $J$.
Push each cycle slightly into the interior of its respective three-ball, keeping the cycles disjoint.
Each cycle then bounds a surface in its own ball, disjoint from the other cycle, so their linking number is zero.
This contradicts the Hopf linking of the two cycles.
\end{proof}

\paragraph{Summing the internal link variables.}
First suppose $r>0$, so there are no internal variables on the $J$ side.
Set $X_{\mathrm{ext}}:=\sum_{i\notin R}a_i$ and
$H_{I,\mathrm{ext}}:=\prod_{(i,j)\in I,\,i\notin R}\chi_k(a_i,\at_j)$.
The $r$ internal variables enter only through their sum $u=\sum_{i\in R}a_i$, with $N^{r-1}$ assignments for each value of $u$.
Define $H_{I,\mathrm{int}}:=H_{I} H_{I,\mathrm{ext}}^{-1}$. Using Eq.~\eqref{symbilin}, we can simplify this expression as follows:
\begin{align}
  H_{I,\mathrm{int}}
  =\prod_{i \in R} \prod_{j=1}^{4}\chi_k(a_i,\at_j)
  =\prod_{i \in R} \chi_k(a_i,Y)
  =\chi_k(u,Y).
\end{align}
Using $X=X_{\mathrm{ext}}+u$ and Eq.~\eqref{eq:defect-forward-fourier}, we obtain
\begin{align}
  \sum_{\{a_i\in\Zb_N\}_{i\in R}}w(X)H_I
  &=N^{r-1}H_{I,\mathrm{ext}}
    \sum_{u\in\Zb_N}w(X_{\mathrm{ext}}+u)\chi_k(u,Y)\notag\\
  &=N^{r-1/2}\wt(Y)H_{I,\mathrm{ext}}\chi_k(-X_{\mathrm{ext}},Y)\notag\\
  &=N^{r-1/2}\wt(Y)H_J^{-1}.
  \label{eq:defect-row-sum}
\end{align}
In the last step, we used $H_{I,\mathrm{ext}}H_J=\chi_k(X_{\mathrm{ext}},Y)$, which follows from Eq.~\eqref{eq:defect-complementary-factors} after removing the complete rows.

For $c>0$, the same argument in the reverse direction uses Eq.~\eqref{eq:defect-reverse-fourier}.
Writing $Y_{\mathrm{ext}}:=\sum_{j\notin C}\at_j$ and
$H_{J,\mathrm{ext}}:=\prod_{(i,j)\in J,\,j\notin C}\chi_k(a_i,\at_j)$, we obtain
\begin{align}
  \sum_{\{\at_j\in\Zb_N\}_{j\in C}}\wt(Y)H_J^{-1}
  &=N^{c-1}H_{J,\mathrm{ext}}^{-1}
    \sum_{v\in\Zb_N}\wt(Y_{\mathrm{ext}}+v)\chi_k(-X,v)\notag\\
  &=N^{c-1/2}w(X)H_{J,\mathrm{ext}}^{-1}\chi_k(X,Y_{\mathrm{ext}})\notag\\
  &=N^{c-1/2}w(X)H_I.
  \label{eq:defect-column-sum}
\end{align}
Lemma~\ref{lemma:defect-ball-move} ensures that these two cases exhaust all allowed moves.

\paragraph{Matching the site and link weights.}
Let $\mathcal N_I$ and $\mathcal N_J$ denote the products of constant site and link weights before and after the move, with unchanged factors outside the cell suppressed.
For $r>0$, all four dual links and their four sites are active on both sides, because every column meets $I$.
The original plaquette initially has four active links and four active sites.
After the move, its surviving active links are precisely the $4-r$ rows that meet $J$.
The tetrahedra of the three-ball $J$ are connected through triangular faces.
Two tetrahedra sharing such a face have either the same row or adjacent rows in $p$; hence the rows meeting $J$ form a connected subset of the row cycle.
Since $1\leq r\leq3$, the surviving original links form a nonempty path with $4-r$ links and $5-r$ sites.
Thus, $r$ active links and $r-1$ active sites disappear, and the inactive weights are one.
Consequently,
\begin{align}
  \frac{\mathcal N_I}{\mathcal N_J}
  =s^{r-1}l^r=N^{-r+1/2}.
  \label{eq:defect-row-normalization}
\end{align}
This exactly cancels the factor $N^{r-1/2}$ in Eq.~\eqref{eq:defect-row-sum}.
For $c>0$, apply the same counting to the reverse move, exchanging the two cycles and the two patches.
It gives $\mathcal N_J/\mathcal N_I=s^{c-1}l^c=N^{-c+1/2}$, which cancels the factor in Eq.~\eqref{eq:defect-column-sum}.
In either case, we have the defect commutation relation
\begin{align}
  \mathcal N_I\sum_{\{a_i\in\Zb_N\}_{i\in R}}w(X)H_I
  =\mathcal N_J\sum_{\{\at_j\in\Zb_N\}_{j\in C}}\wt(Y)H_J^{-1},
  \label{eq:defect-commutation-relation}
\end{align}
where a sum over an empty set of variables leaves the summand unchanged.
This equality holds for every fixed assignment of external active link variables.
Multiplying by the unchanged weights outside the cell and summing over those variables proves invariance of the full partition function under the allowed local deformation, and hence under any sequence of such moves.
No positivity or self-duality assumption was used.
When $w=\wt$, the same proof establishes local topological invariance of the duality defect within a single theory.

\subsection{\texorpdfstring{$S^3$}{S3} expectation value and connected sum}
\label{sec:defect-sphere-connected-sum}

We now consider two local operations that change the topology of the interface: removing a spherical component and joining two components by a tube.
Their coefficients need not be one, since these operations are not among the topology-preserving moves of the previous subsection.
These relations are independent of the details of the dynamics and are determined solely by the structure of the ``symmetry.''
In this sense, they are analogs of Ward--Takahashi identities.
We denote the interface on an oriented closed three-manifold $M$ by $\Ic(M)$.
The orientation of $M$ specifies which side carries $w$ and which carries $\wt$.
We use $\expval{\cdots}$ to denote unnormalized expectation values, computed with the site and link weights in Eq.~\eqref{dualitydefectweight}.
In particular, $\expval{1}$ is the partition function without operator insertions.

\paragraph{Spherical interface.}
Place an interface on all sixteen facets of a single 16-cell, with weight $w$ inside and $\wt$ outside.
The interface is a three-sphere, and the product of its tetrahedral weights is
\begin{align}
  H_U=\prod_{i,j=1}^4\chi_k(a_i,\at_j)=\chi_k(X,Y),
\end{align}
where $X$ and $Y$ are defined in Eq.~\eqref{eq:defect-patch-factors}.
All four original link variables are internal, while the four dual link variables are held fixed.
For each value of $X$, there are $N^3$ assignments of $a_1,\ldots,a_4$.
Consequently,
\begin{align}
  \sum_{a_1,\ldots,a_4\in\Zb_N}w(X)H_U
  &=N^3\sum_{X\in\Zb_N}w(X)\chi_k(X,Y)\notag\\
  &=N^{7/2}\wt(Y).
  \label{eq:defect-sphere-sum}
\end{align}
Before we remove the sphere, the two plaquette cycles together have eight active sites and eight active links.
Once we remove the sphere and fill the cell with the exterior theory, only the four dual sites and four dual links remain active.
Including their weights gives the local removal identity
\begin{align}
  N^{-8}\sum_{a_1,\ldots,a_4\in\Zb_N}w(X)H_U
  =\frac{1}{\sqrt N}\,N^{-4}\wt(Y).
  \label{eq:defect-sphere-removal}
\end{align}
Thus, an empty spherical interface that can be reduced to this configuration by the allowed local deformations satisfies
\begin{align}
  \expval{\Ic(S^3)\cdots}
  =\frac{1}{\sqrt N}\expval{\cdots},
  \label{eq:defect-sphere-expectation}
\end{align}
where $\cdots$ denotes arbitrary operator insertions outside the sphere, held fixed on both sides of the identity.
On the right-hand side, the interior of the sphere is filled with the exterior theory.
The same coefficient follows from the inverse Fourier transform when the two theories are exchanged and the interface orientation is reversed.
It is independent of $k$ and of the choice of the dual pair $(w,\wt)$.
For $N=2$, this reproduces the expectation value of the duality defect on $S^3$ obtained in Ref.~\cite{Koide:2021zxj}.

\begin{figure}[htbp]
  \centering
  \begin{tikzpicture}[font=\small]
    \begin{scope}[xshift=1.8cm,yshift=1.3cm,scale=0.65]
      \fill[gray!10] (0,0) rectangle (4,-4);
      \fill[green!25] (0,0) rectangle (1,-1);
      \fill[green!25] (2,-2) rectangle (3,-3);
      \draw[step=1,gray] (0,0) grid (4,-4);
      \foreach \i in {1,...,4}{
        \node[left] at (0,-\i+0.5) {$\ell_\i$};
        \node[above] at (\i-0.5,0) {$\ltl_\i$};
      }
      \node[below] at (2,-4.4) {$I=\{(1,1), (3,3)\}$};
    \end{scope}
  \end{tikzpicture}
  \caption{The configuration used for the connected sum.  $I$ consists of two disjoint tetrahedra.}
  \label{fig:two-balls}
\end{figure}
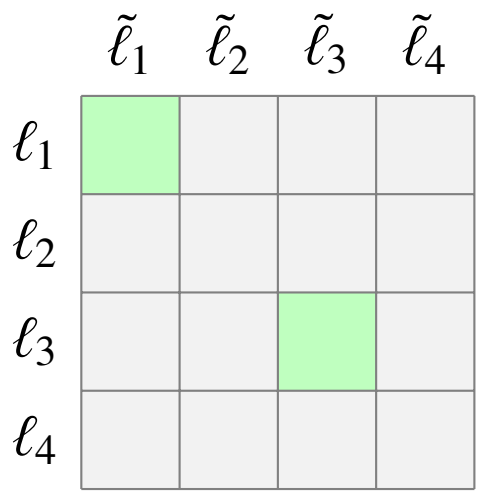

\paragraph{Connected sum.}
Consider the configuration shown in Fig.~\ref{fig:two-balls}.
The two tetrahedra in $I$ have no common vertices and form two disjoint three-balls.
Their complement $J$ in the boundary three-sphere is a cylinder $S^2\times[0,1]$.
Replacing $I$ by $J$ therefore removes the interiors of two three-balls from the interface and joins their boundary two-spheres by a tube.
If the balls lie on distinct connected components $M$ and $M'$, the resulting interface is supported on the connected sum $M\#M'$.

The Fourier calculation of Eq.~\eqref{eq:defect-column-sum} applies even though this replacement changes the topology.
There are two internal dual links, $\at_2$ and $\at_4$, so that
\begin{align}
  \sum_{\at_2,\at_4\in\Zb_N}\wt(Y)H_J^{-1}
  =N^{3/2}w(X)H_I,\qquad
  H_I=\chi_k(a_1,\at_1)\chi_k(a_3,\at_3).
  \label{eq:defect-connected-sum-fourier}
\end{align}
Here all $a_i$, as well as $\at_1$ and $\at_3$, are fixed external variables.
On the $I$ side, the original cycle has four active links and four active sites, while the dual lattice contributes two opposite active links and their four endpoints.
On the $J$ side, both cycles are fully active.
The corresponding constant factors are therefore
\begin{align}
  \mathcal N_I=s^8l^6=N^{-7},\qquad
  \mathcal N_J=s^8l^8=N^{-8}.
\end{align}
Combining these factors with Eq.~\eqref{eq:defect-connected-sum-fourier}, we obtain
\begin{align}
  \mathcal N_J\sum_{\at_2,\at_4\in\Zb_N}\wt(Y)H_J^{-1}
  =\sqrt N\,\mathcal N_I w(X)H_I.
  \label{eq:defect-connected-sum-local}
\end{align}
Exchanging the two theories and reversing the interface orientation gives the same coefficient for the opposite assignment of region weights.

Since Eq.~\eqref{eq:defect-connected-sum-local} holds for every assignment of external variables, it remains valid after attaching the unchanged exterior and summing over those variables.
For a standard tube with compatible region labels and no other operator insertions in the surgery region, it gives
\begin{align}
  \expval{\Ic(M\#M')\cdots}
  =\sqrt N\,
  \expval{\Ic(M)\Ic(M')\cdots},
  \label{eq:defect-connected-sum}
\end{align}
where $\cdots$ denotes arbitrary additional operator insertions outside the surgery region, held fixed on both sides.
As a check, removing two empty spheres independently multiplies the remaining correlator by $N^{-1}$.
Joining them first gives the total factor $\sqrt N\,N^{-1}=N^{-1/2}$, as required by $S^3\#S^3\cong S^3$.

The same local operation applied to two balls on one connected component attaches a handle, replacing $M$ by $M\#(S^2\times S^1)$.
Each such handle multiplies the correlator by $\sqrt N$.
Starting from an empty standard sphere and attaching $g$ standard handles gives an interface supported on
\begin{align}
  M_g:=\mathop{\#}_{j=1}^{g}(S^2\times S^1),\qquad M_0:=S^3.
\end{align}
Combining the handle factors with Eq.~\eqref{eq:defect-sphere-expectation}, we obtain
\begin{align}
  \expval{\Ic(M_g)\cdots}=N^{(g-1)/2}\expval{\cdots},
  \label{eq:defect-handle-expectation}
\end{align}
where the additional insertions lie outside a four-ball containing the interface and the entire sequence of local deformations used to remove it.
In particular, removing a standard $S^2\times S^1$ interface under these conditions leaves the correlator unchanged.
These identities apply to the embeddings obtained by the stated local operations.
At self-duality, $w=\wt$, all of these relations become relations for duality defects within a single theory.

\section{Symmetry defects and junctions}
\label{sec:symmetry-junctions}

In this section, we construct center symmetry defects and their junctions with the duality interface.  Although we have already described the center symmetry using a background gauge field, it is convenient to define the defects through local weights, as in \cite{Aasen:2016dop,Koide:2021zxj}, in order to describe junctions and crossing relations.
We then derive crossing relations from their local weights.
Throughout this section, the bulk weights are an even dual pair $w,\wt=\Fs_k w$, and the site and link weights are those in Eq.~\eqref{dualitydefectweight}.
When $w=\wt$, the interfaces become defects in a single theory.

\subsection{Center symmetry defect}
\label{sec:center-defect}

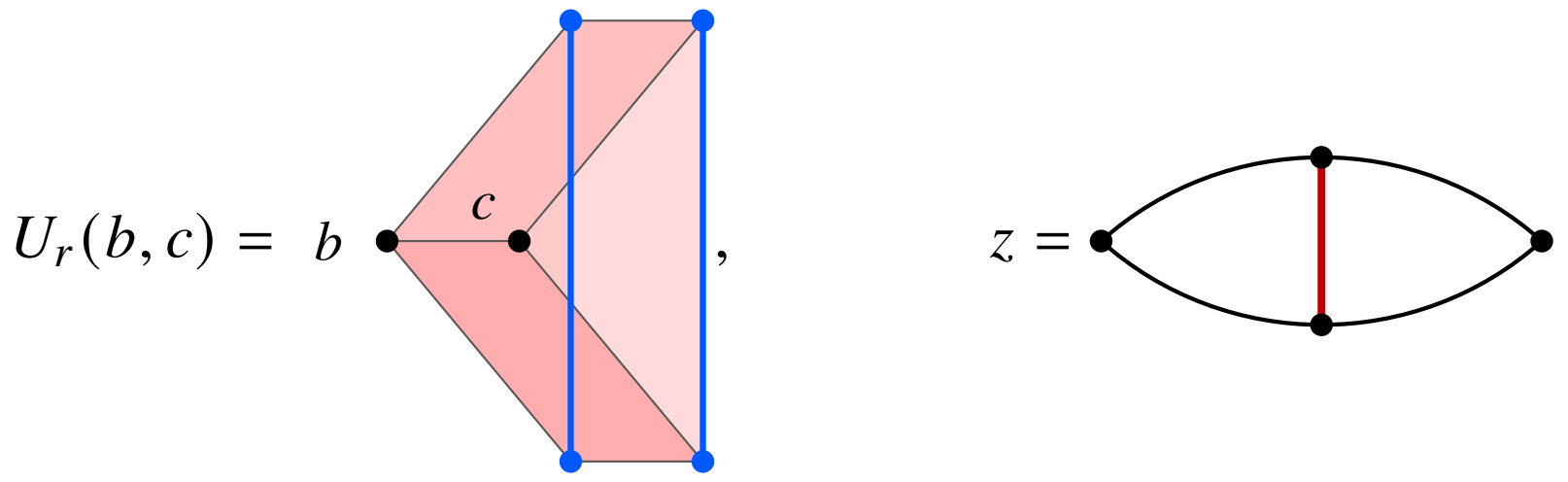
\begin{figure}[htbp]
  \centering
  $U_r(b,c)=\vcenter{\hbox{%
  \begin{tikzpicture}[font=\small,line join=round]
    \coordinate (mb) at (3.85,0);
    \coordinate (ub) at (5.1,1.5);
    \coordinate (vb) at (5.1,-1.5);
    \coordinate (mc) at (4.75,0);
    \coordinate (uc) at (6,1.5);
    \coordinate (vc) at (6,-1.5);
    \fill[red!18] (mb)--(ub)--(vb)--cycle;
    \fill[red!25] (mb)--(ub)--(uc)--(mc)--cycle;
    \fill[red!32] (mb)--(vb)--(vc)--(mc)--cycle;
    \fill[red!22,fill opacity=0.65] (mc)--(uc)--(vc)--cycle;
    \draw[black!65] (ub)--(mb)--(vb) (uc)--(mc)--(vc);
    \draw[black!65] (ub)--(uc) (vb)--(vc) (mb)--(mc);
    \draw[blue!65!cyan,line width=1.2pt] (ub)--(vb) (uc)--(vc);
    \foreach \p in {ub,vb,uc,vc}
      \fill[blue!65!cyan] (\p) circle (2.2pt);
    \fill (mb) circle (2.2pt) node[left=4pt] {$b$};
    \fill (mc) circle (2.2pt) node[above left] {$c$};
  \end{tikzpicture}%
  }},\qquad\qquad z=\vcenter{\hbox{%
  \begin{tikzpicture}[font=\small,line cap=round,line join=round]
    \draw[red!75!black,line width=1.5pt] (1.5,-0.57)--(1.5,0.57);
    \draw[thick] (0,0)..controls (0.85,0.76) and (2.15,0.76)..(3,0);
    \draw[thick] (0,0)..controls (0.85,-0.76) and (2.15,-0.76)..(3,0);
    \foreach \p in {(0,0),(3,0),(1.5,0.57),(1.5,-0.57)}
      \fill \p circle (2.2pt);
  \end{tikzpicture}%
  }}$
  \caption{Local building blocks of the center symmetry defect.
  Left: a triangular prism whose black dots mark active-link midpoints,
  while blue dots and blue edges denote inactive sites and links.
  The two active-link copies carry variables $b$ and $c$ in the same reference orientation; their common endpoints are not shown.
  Right: the doubled active links share their endpoints, and the red segment joins their midpoints.}
  \label{fig:center-defect-prism}
\end{figure}

\paragraph{Geometry.}
The defect generating the one-form center symmetry is supported on an oriented two-dimensional surface $\Sigma$ and is labeled by $r\in\Zb_N$.
We use the triangular-prism realization of Ref.~\cite{Koide:2021zxj}.
A triangle joins an inactive link to the midpoint of an adjacent active link.
Gluing such triangles gives $\Sigma$.
Along this surface, we double the inactive links and their endpoints, and double each active link whose midpoint lies on the surface while retaining its endpoints.
Each triangle consequently becomes a triangular prism, as illustrated in Fig.~\ref{fig:center-defect-prism}.
The two active copies carry variables $b,c\in\Zb_N$, with the same reference orientation.
Plaquettes use the copy on their respective side of the cut.

\paragraph{Weights.}
Choose the orientation of a prism so that crossing it changes the active link variable from $b$ to $c=b+r$.
Its weight, together with the compensating factor for each doubled active link, is
\begin{align}
  U_r(b,c):=\delta_{c,b+r},\qquad z:=\sqrt N.
  \label{eq:center-local-weight}
\end{align}
Reversing the orientation replaces $r$ by $-r$.
If several prisms share the same pair of active links, they impose the same constraint, while $z$ is assigned only once to that pair.
Since $zl=1$, the extra active link weight is canceled.
The doubled inactive elements have weight one, and no extra active site is introduced.

\paragraph{Topological invariance.}
Consider two consecutive active links $1,2$ around a plaquette in a 16-cell.
Joining their midpoints to each of the four inactive links gives the eight triangular faces of an octahedron, as in Ref.~\cite{Koide:2021zxj}.
A local move replaces a disk of faces $H$ by the complementary disk $H^c$ with the same oriented boundary, keeping the surface and link variables outside the octahedron fixed.

First take $H$ to be the four faces meeting the midpoint of link $1$.
Before the move the surface splits link $1$, and afterwards it splits link $2$.
Orient the active links along the plaquette boundary, so that link $1$ enters their common vertex and link $2$ leaves it.
Choose the surface orientation so that the change from copy $b_m$ to copy $c_m$ is $-r$ on link $1$ and $+r$ on link $2$.
The external copies are $c_1$ and $b_2$. Write $A=a_3+a_4$ for the contribution of the other two links.
The local weights then satisfy
\begin{align}
  \underbrace{(zl)\sum_{b_1\in\Zb_N} U_{-r}(b_1,c_1)^4\,w(b_1+b_2+A)}_{\text{before (link }1\text{ split)}}
  =
  \underbrace{(zl)\sum_{c_2\in\Zb_N} U_r(b_2,c_2)^4\,w(c_1+c_2+A)}_{\text{after (link }2\text{ split)}} .
  \label{eq:center-deformation}
\end{align}
The Kronecker delta function from the weight $U$ fixes $b_1=c_1+r$ on the left and $c_2=b_2+r$ on the right.
Both sides therefore give $w(c_1+b_2+A+r)$ for every choice of external links.
The fourth power counts the four prisms sharing the split link. They impose the same delta function, and the single extra link contributes $zl=1$.

For a general patch, label a face by $(m,j)$, where $m=1,2$ specifies the active midpoint and $j=1,\ldots,4$ the inactive link.
The plaquette uses the $b$ copies before the move and the $c$ copies afterwards, giving fluxes $f_b=b_1+b_2+A$ and $f_c=c_1+c_2+A$.
With $\epsilon_1=-1$ and $\epsilon_2=+1$, the oriented $\Zb_N$ version of the commutation relation is
\begin{align}
  \underbrace{\sum_{\text{internal }b} w(f_b)
    \prod_{(m,j)\in H} U_{\epsilon_m r}(b_m,c_m)}_{\text{surface on }H}
  =
  \underbrace{\sum_{\text{internal }c} w(f_c)
    \prod_{(m,j)\in H^c} U_{\epsilon_m r}(b_m,c_m)}_{\text{surface on }H^c}.
  \label{eq:center-patch-deformation}
\end{align}
If all four faces at midpoint $m$ lie in $H$, only the left side sums over $b_m$. If they all lie in $H^c$, only the right side sums over $c_m$.
In either case the sum fixes the extra copy uniquely.
If both patches meet the midpoint, both copies are external and both sides impose the same constraint.
Thus either both weights vanish or $c_1=b_1-r$ and $c_2=b_2+r$, giving $f_c=f_b$.
Taking $H$ to be the entire octahedral surface gives the empty configuration on the right, with coefficient one.
Successive moves therefore preserve the weight of a closed surface and allow a contractible surface to be removed when the deformation region contains no other insertions.
In the background-field description of Eq.~\eqref{PartitionFunction}, the same invariance follows because a local deformation changes $B$ by $\dd\lambda^1$.
The change of summation variable $a\mapsto a+\lambda^1$ then preserves every plaquette flux $\dd a-B$.

\subsection{Junctions}
\label{sec:defect-junctions}

\begin{figure}[htbp]
  \centering
  \begin{tikzpicture}[font=\small,line cap=round,line join=round]
    \definecolor{interfacegreen}{RGB}{0,150,35}
    \begin{scope}
      \coordinate (atop) at (0,1.25);
      \coordinate (abot) at (0,-1.25);
      \coordinate (btop) at (1.65,0.55);
      \coordinate (bbot) at (1.65,-1.95);
      \coordinate (ctop) at (2.7,1.35);
      \coordinate (cbot) at (2.7,-1.15);
      \fill[interfacegreen!28] (atop)--(ctop)--(cbot)--(abot)--cycle;
      \fill[interfacegreen!42] (atop)--(btop)--(bbot)--(abot)--cycle;
      \fill[interfacegreen!32] (abot)--(bbot)--(cbot)--cycle;
      \fill[red!35] (btop)--(ctop)--(cbot)--(bbot)--cycle;
      \fill[interfacegreen!22] (atop)--(btop)--(ctop)--cycle;
      \draw[interfacegreen!80!black] (atop)--(btop)--(ctop)--cycle;
      \draw[interfacegreen!80!black] (abot)--(bbot)--(cbot)--cycle;
      \draw[black,line width=1.2pt] (atop)--(abot)
        node[midway,left=5pt] {$a$};
      \draw[blue!65!cyan,line width=1.2pt] (btop)--(bbot) (ctop)--(cbot);
      \foreach \p in {atop,abot}
        \fill (\p) circle (2.2pt);
      \foreach \p in {btop,bbot,ctop,cbot}
        \fill[blue!65!cyan] (\p) circle (2.2pt);
      \node at (1.35,-2.4) {(a) $J_r(a)$};
    \end{scope}

    \begin{scope}[xshift=6.2cm,yshift=-0.2cm]
      \coordinate (ltop) at (0,1.35);
      \coordinate (lbot) at (0,-1.15);
      \coordinate (rtop) at (2.8,0.95);
      \coordinate (rbot) at (2.8,-1.55);
      \coordinate (bmid) at (2.25,-0.4);
      \coordinate (cmid) at (3.35,-0.2);
      \fill[interfacegreen!35] (ltop)--(rtop)
        ..controls (3.15,0.65) and (3.35,0.2)..(cmid)
        ..controls (3.35,-0.7) and (3.15,-1.2)..(rbot)
        --(lbot)--cycle;
      \draw[interfacegreen!80!black] (ltop)--(rtop) (lbot)--(rbot);
      \draw[blue!65!cyan,line width=1.2pt] (ltop)--(lbot);
      \draw[red!75!black,line width=1.8pt] (bmid)--(cmid);
      \draw[black,line width=1.2pt] (rtop)
        ..controls (2.45,0.45) and (2.25,0)..(bmid)
        ..controls (2.25,-0.9) and (2.45,-1.3)..(rbot);
      \draw[black,line width=1.2pt] (rtop)
        ..controls (3.15,0.65) and (3.35,0.2)..(cmid)
        ..controls (3.35,-0.7) and (3.15,-1.2)..(rbot);
      \foreach \p in {rtop,rbot,bmid,cmid}
        \fill (\p) circle (2.2pt);
      \foreach \p in {ltop,lbot}
        \fill[blue!65!cyan] (\p) circle (2.2pt);
      \node[left=4pt] at (bmid) {$b$};
      \node[right=4pt] at (cmid) {$c$};
      \node at (1.65,-2.2) {(b) $\Jt_r(b,c)$};
    \end{scope}
  \end{tikzpicture}
  \caption{Schematic picture of local junctions of a center symmetry defect with the duality interface.
  Green and red indicate the interface and symmetry defect, respectively;
  black and blue edges denote active and inactive links.
  (a) The shared link is inactive on the symmetry-defect side,
  and the corresponding active link on the other side carries $a$.
  (b) The shared active link splits into copies carrying $b,c$, with their endpoints retained.
  The junction weights use the standard orientations of Eq.~\eqref{eq:junction-weights}.}
  \label{fig:defect-junctions}
\end{figure}
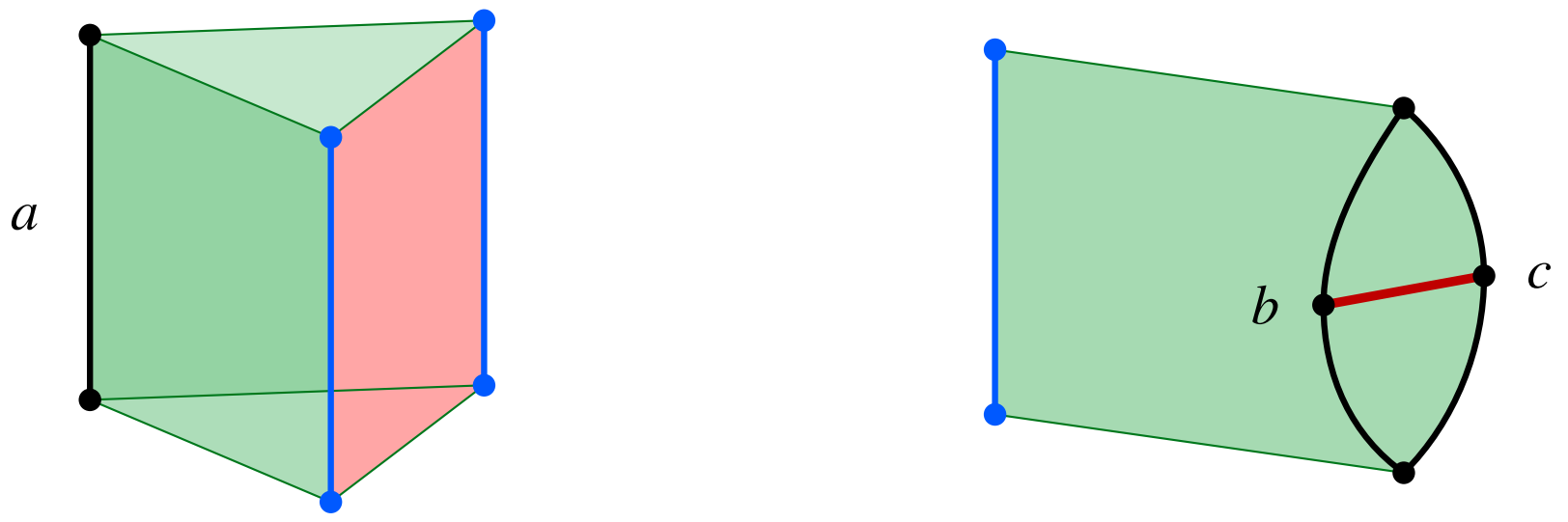

\paragraph{Geometry and weights.}
A center symmetry defect can end on the duality interface $M$.
Its boundary $\partial\Sigma\subset M$ is a one-dimensional junction.
There are two local descriptions, according to whether the shared link is inactive or active on the side containing the symmetry defect.
They are illustrated in Fig.~\ref{fig:defect-junctions}.
In the first case, the corresponding link on the other side of the interface is active and carries a variable $a$.
In the second case, the symmetry defect splits the active link into copies $b,c$.
For the standard orientations, we assign
\begin{align}
  J_r(a):=\chi_k(r,a),\qquad
  \Jt_r(b,c):=\delta_{c,b+r}.
  \label{eq:junction-weights}
\end{align}
The delta function at a $\Jt_r$ junction is the continuation of the symmetry-defect constraint.
The factor $z$ is still counted once per doubled active link, including a link at the junction.

\paragraph{Gauge invariance.}
The weight $\Jt_r$ is invariant under the simultaneous gauge transformation of its two parallel links.
Under $a\mapsto a+\lambda_j-\lambda_i$, an individual junction weight transforms as $J_r(a+\lambda_j-\lambda_i)=J_r(a)J_r(\lambda_j-\lambda_i)$.
If the links carrying $J_r$ form a closed oriented chain, these phases cancel between consecutive links. At a meeting of such chains the phases cancel when the sum of incoming labels equals the sum of outgoing labels.
For a closed junction containing both $J_r$ and $\Jt_r$, the links carrying $J_r$ can form an open path.
In that case the adjacent interface prisms evaluated on the shifted link copies supply the missing endpoint phases, as follows from Eq.~\eqref{eq:junction-local-identities} below. The center symmetry defect supported on a disk in Eq.~\eqref{eq:solid-torus-relation} gives an explicit example.

\paragraph{Topological invariance.}
First move the boundary of the symmetry defect across a single interface prism, keeping the external link variables $a$ and $\at$ fixed.
Before the move, the original link is split, as in Fig.~\ref{fig:defect-junctions}(b), and the copy $b$ enters the interface weight $D(b,\at)$.
After the move, the prism carries $D(a,\at)$ and the junction lies on the dual link $\at$, as in Fig.~\ref{fig:defect-junctions}(a).
For the standard orientations the two local weights obey
\begin{align}
  \underbrace{(zl)\sum_{b\in\Zb_N}\Jt_r(a,b)\,D(b,\at)}_{\text{before}}
  =D(a+r,\at)
  =\underbrace{D(a,\at)\,J_r(\at)}_{\text{after}}.
  \label{eq:junction-local-identities}
\end{align}
The first equality follows because $\Jt_r(a,b)=\delta_{b,a+r}$ selects one value of $b$ and the extra link has net weight $zl=1$.
The second follows from the multiplicativity of $D$ and the definition of $J_r$ in Eq.~\eqref{eq:junction-weights}.
Thus the factor $J_r(\at)$ on the dual link accounts exactly for the shift of the original link.
For a surface incident from the dual side, exchanging $a$ and $\at$ gives the same identity with $J_r(a)$ on the original link.

Now slide the junction along a fixed interface patch $I$ in a region free of other insertions.
Before the slide the surface splits original link $i$, and afterwards it splits the adjacent link $h$.
Use the plaquette flux $X$ and the interface weight $H_I$ of Eq.~\eqref{eq:defect-patch-factors}.
If the surface splits link $s$, the sum over its extra copy sets that copy to $a_s+r$.
The plaquette then has flux $X+r$, and each prism $(s,j)\in I$ contributes one factor $J_r(\at_j)$ by Eq.~\eqref{eq:junction-local-identities}.
Writing $a=(a_1,\ldots,a_4)$, $\at=(\at_1,\ldots,\at_4)$, and $e_s$ for the unit vector in the $s$th link variable, we obtain
\begin{align}
  H_I(a+r e_s,\at)=H_I(a,\at)\,J_r(Y_s),
  \qquad Y_s:=\sum_{j:(s,j)\in I}\at_j .
\end{align}
Here $J_r(Y_s)=\prod_{j:(s,j)\in I}J_r(\at_j)$ is the product of the junction weights on these links.
Omitting the factors outside the move, we can now equate the two local weights,
\begin{align}
  \underbrace{w(X+r)\,H_I(a+r e_i,\at)}_{\text{before (link }i\text{ split)}}
  =
  \underbrace{w(X+r)\,H_I(a+r e_h,\at)\,J_r(Y_i)J_{-r}(Y_h)}_{\text{after (link }h\text{ split)}} .
  \label{eq:junction-patch-deformation}
\end{align}
Moving the split off link $i$ leaves a factor $J_r(\at_j)$ on each dual link in row $i$.
Moving it onto link $h$ uses Eq.~\eqref{eq:junction-local-identities} in reverse and gives $J_{-r}(\at_j)$ on each dual link in row $h$.
The two factors cancel on every dual link shared by the rows, so their product is precisely the weight of the oriented junction segment added by the slide.
Both sides thus equal $w(X+r)H_I(a,\at)J_r(Y_i)$ for each choice of external link variables.
Together with the bulk surface move in Eq.~\eqref{eq:center-deformation}, this proves local invariance of the surface and its boundary on the interface.

Finally, consider a symmetry defect supported on a disk whose boundary can be contracted on the interface within an otherwise empty local patch.
Let $u$ be the internal link variable, $X_{\mathrm{ext}}$ the flux from the other original links, and $Y$ the oriented sum of the dual links around the disk boundary.
The product of interface weights in this row is $\prod_jD(u,\at_j)=D(u,Y)$.
The disk and its boundary can be removed by the local identity
\begin{align}
  \underbrace{\sum_{u\in\Zb_N}w(X_{\mathrm{ext}}+u+r)D(u,Y)J_r(Y)}_{\text{with disk and junction}}
  =\underbrace{\sum_{v\in\Zb_N}w(X_{\mathrm{ext}}+v)D(v,Y)}_{\text{after removal}} .
  \label{eq:junction-disk-removal}
\end{align}
The product of boundary weights is $J_r(Y)=\prod_j J_r(\at_j)$ in the chosen orientations.
Applying Eq.~\eqref{eq:junction-local-identities} to each prism gives $D(u,Y)J_r(Y)=D(u+r,Y)$, so the bijective change of variable $v=u+r$ gives the right-hand side with coefficient one.

\subsection{Crossing relations}
\label{sec:defect-crossing}

\begin{figure}[htbp]
  \centering
  \includegraphics[width=10cm]{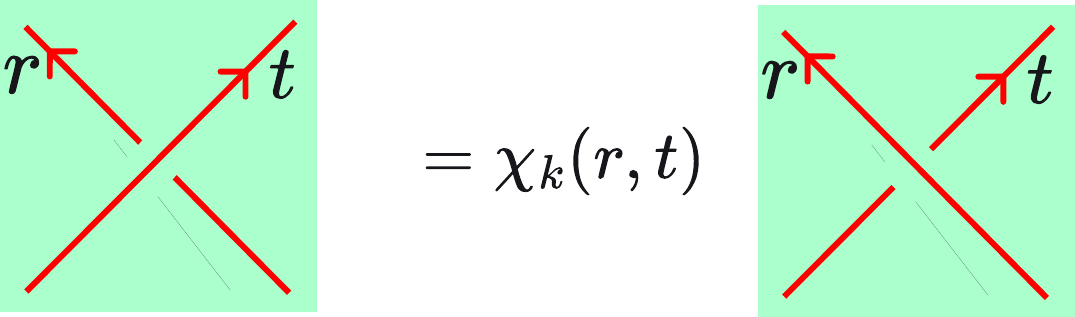}
  \caption{Braiding in the three-dimensional worldvolume of the duality interface.
  The red lines mark the boundaries of center symmetry defects that meet the interface from opposite sides.}
  \label{fig:brading}
\end{figure}

\paragraph{Braiding on the duality interface.}
The junctions inherited from the two sides of the interface have a nontrivial mutual crossing phase.
For the crossing in which a $J_r$ factor is moved from the $b$ copy of a link to the $c$ copy constrained by $\Jt_t$, we have
\begin{align}
  J_r(c)\Jt_t(b,c)
  =\chi_k(r,t)\Jt_t(b,c)J_r(b).
  \label{eq:junction-braiding}
\end{align}
Indeed, on the support of the delta function, $c=b+t$, so that
$J_r(c)=\chi_k(r,t)J_r(b)$.
Here $\chi_k(r,t)$ is the scalar braiding coefficient: it depends only on the surface labels $r,t$, whereas $J_r$ and $\Jt_t$ are the local junction weights.
We take this to be the positive crossing convention.
Geometrically, this is the braiding of the boundary lines of symmetry defects attached from opposite sides of the three-dimensional interface as shown in Fig.~\ref{fig:brading}.
This braiding phase agrees with that obtained in
Ref.~\cite{Antinucci:2023ezl}.
The attached surfaces are retained in this relation.
Within either family, the local junction factors commute: characters multiply and shifts add in $\Zb_N$.

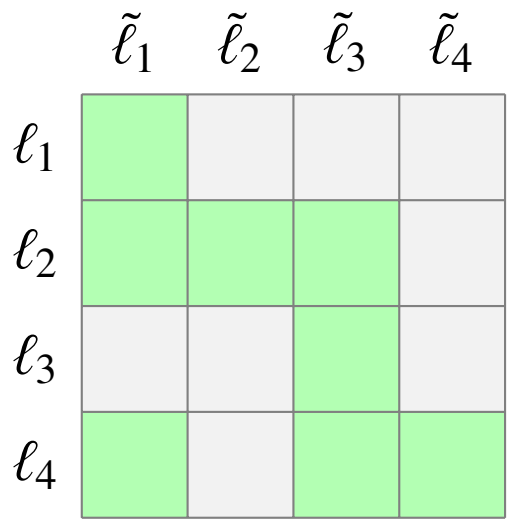
\begin{figure}[htbp]
  \centering
  \begin{tikzpicture}[font=\small,scale=0.7]
    \fill[gray!10] (0,0) rectangle (4,-4);
    \foreach \i/\j in {0/0,1/0,1/1,1/2,2/2,3/0,3/2,3/3}
      {\fill[green!30] (\j,-\i) rectangle ++(1,-1);}
    \draw[step=1,gray] (0,0) grid (4,-4);
    \foreach \i in {1,...,4}{
      \node[left] at (0,-\i+0.5) {$\ell_\i$};
      \node[above] at (\i-0.5,0) {$\ltl_\i$};
    }
    \node[anchor=west,align=left] at (4.5,-1.4)
      {};
  \end{tikzpicture}
  \caption{Complementary solid-torus patches for Eq.~\eqref{eq:solid-torus-relation}.
  Each entry represents a tetrahedron: a green one is in $I$ and a gray one is in $J$.}
  \label{fig:solid-torus-patches}
\end{figure}

\paragraph{Solid torus relation.}
We next replace a solid-torus patch of the interface by its complementary solid torus in the boundary of a 16-cell.
Unlike the three-ball moves of Sec.~\ref{sec:defect-topological-invariance}, this operation exchanges the two cycles of their common boundary torus and requires a sum over symmetry-defect insertions.
One concrete choice is
\begin{align}
  I=\{(1,1),(2,1),(2,2),(2,3),(3,3),(4,1),(4,3),(4,4)\},
  \qquad J=U\setminus I.
  \label{eq:solid-torus-patches}
\end{align}
The two patches are shown in Fig.~\ref{fig:solid-torus-patches}.
Each is a regular neighborhood of a cycle and hence a solid torus; their common boundary is a two-torus.
Every row and every column meets both patches, so there are no internal active links to sum over on either side.

Using $X,Y,H_I,H_J$ as defined in Eq.~\eqref{eq:defect-patch-factors}, we obtain the local crossing relation
\begin{align}
  w(X)H_I(a,\at)
  =\frac{1}{\sqrt N}\sum_{t\in\Zb_N}
    \wt(Y+t)H_J(a,\at+t e_1)^{-1}
    \prod_{i\in\{1,2,4\}}J_{-t}(a_i).
  \label{eq:solid-torus-relation}
\end{align}
Here $e_1$ shifts the first dual link variable.
In the term labeled by $t$, a center symmetry defect supported on a disk splits the dual link $j=1$ and shifts the dual flux by $+t$.
In this column, the triangles that make up the disk end on original links $1,2,4$, giving the three junction weights displayed in Eq.~\eqref{eq:solid-torus-relation}.
The $\Jt_t$ constraints set the shifted copy to $\at_1+t$.
The only tetrahedron in this column of $J$ is $(3,1)$, so the shifted interface weight is
\begin{align}
  H_J(a,\at+t e_1)^{-1}
  =H_J(a,\at)^{-1}
    \frac{D(a_3,\at_1+t)^{-1}}{D(a_3,\at_1)^{-1}}.
\end{align}
The signs follow from the opposite orientation of the $J$ patch.
Thus the shifted interface weight and the junction weights are kept separate.
Their combined ratio to $H_J(a,\at)^{-1}$ depends only on the original plaquette flux $X$ and is gauge invariant.
The sum over $t$ runs over the surface labels.

For completeness, the relation follows directly from the inverse Fourier transform.
Multiplicativity gives
\begin{align}
  H_I(a,\at+t e_1)\prod_{i\in\{1,2,4\}}J_{-t}(a_i)
  =H_I(a,\at).
\end{align}
Together with Eq.~\eqref{eq:defect-complementary-factors}, this rewrites the interface and junction factors on the right-hand side of Eq.~\eqref{eq:solid-torus-relation} as $H_I(a,\at)D(X,Y+t)^{-1}$.
Here $D(X,Y+t)=H_I(a,\at+t e_1)H_J(a,\at+t e_1)$ is the product of all sixteen interface weights with the shifted dual link.
Setting $h=Y+t$, the right-hand side therefore becomes
\begin{align}
  \frac{H_I(a,\at)}{\sqrt N}\sum_{h\in\Zb_N}\wt(h)D(X,h)^{-1}
  &=H_I(a,\at)\sum_{h\in\Zb_N}(\Fs_{-k})_{Xh}\wt(h)\notag\\
  &=H_I(a,\at)w(X),
\end{align}
where $(\Fs_{-k})_{Xh}$ is the inverse Fourier matrix coefficient and $\Fs_{-k}\wt=w$.
Both configurations have eight active sites and eight active links, giving the common factor
$s^8l^8=N^{-8}$.
Any doubled links on the disk are compensated by $z$.
Consequently, the coefficient $N^{-1/2}$ in Eq.~\eqref{eq:solid-torus-relation} is also the coefficient in the full partition-function identity.

The charged version makes the dependence on the bicharacter explicit.
Define local amplitudes
\begin{align}
  \mathcal A_r&:=w(X-r)H_I(a-r e_1,\at)
    \prod_{j\in\{2,3,4\}}J_{-r}(\at_j),\notag\\
  \mathcal B_t&:=\wt(Y+t)H_J(a,\at+t e_1)^{-1}
    \prod_{i\in\{1,2,4\}}J_{-t}(a_i).
  \label{eq:charged-solid-torus-amplitudes}
\end{align}
For $\mathcal A_r$, the disk splits original link $i=1$: the interface prism $(1,1)\in I$ uses $a_1-r$, while the disk boundary contributes junction weights on dual links $2,3,4$.
The amplitude $\mathcal B_t$ is the term already described in Eq.~\eqref{eq:solid-torus-relation}.
The defects shift the plaquette flux by $-r$ on the $I$ side and by $+t$ on the $J$ side. These opposite signs specify the orientation convention for the two fillings.
Applying the same inverse Fourier transform at flux $X-r$ and including the displayed interface and junction weights yields
\begin{align}
  \mathcal A_r=\frac{1}{\sqrt N}\sum_{t\in\Zb_N}
      \chi_k(r,t)\mathcal B_t,\qquad
  \mathcal B_t=\frac{1}{\sqrt N}\sum_{r\in\Zb_N}
      \chi_k(-r,t)\mathcal A_r.
  \label{eq:charged-solid-torus-crossing}
\end{align}
The factors $\chi_k(r,t)$ and $\chi_k(-r,t)$ here are scalar crossing coefficients depending only on the surface labels; all interface and junction weights are contained in $\mathcal A_r$ and $\mathcal B_t$.
Thus the crossing matrix is precisely $\Fs_k$, and its entries contain the same bicharacter as the braiding phase in Eq.~\eqref{eq:junction-braiding}.
The inverse relation follows from character orthogonality.
All these local equalities hold for fixed external variables and therefore remain valid after attaching the same exterior configuration and performing the remaining link sums.

\section{Half-space gauging}
\label{sec:half-space-gauging}

In Sec.~\ref{sec:duality-defects}, we construct duality defects using the AMF approach \cite{Aasen:2016dop,Koide:2021zxj}. Here, we construct duality defects using the half-space gauging approach \cite{Choi:2021kmx}.
The calculation uses the same finite Fourier transform as the full-space duality, but the original link variables remain at the boundary and couple to the dual link variables.
We will show that this coupling reproduces the tetrahedral-prism weight introduced in Sec.~\ref{sec:duality-defects}.
We focus on its dependence on the fields, leaving the overall normalization of the gauging sum implicit.

\subsection{Gauging in a region with Dirichlet boundary conditions}

Choose a subcomplex $\Omega\subset\Lambda^*$: whenever a cell belongs to $\Omega$, all its boundary cells also belong to $\Omega$.
Write $\Omega_j=\Omega\cap\Lambda_j^*$.
Let $R_\Omega$ be the union of the closed 16-cells associated with the dual plaquettes in $\Omega_2$.
The cochain calculation below applies to any such subcomplex $\Omega$.
To identify $\partial R_\Omega$ with the interface of Sec.~\ref{sec:defect-geometry}, we require it to be an embedded closed piecewise-linear three-manifold in the periodic lattice.
On the original lattice, define the cochains with corresponding support by
\begin{align}
  C_\Omega^r(\Lambda)
  :=\{u\in C^r(\Lambda)\mid u_\sigma=0
      \text{ if }\sigma^*\notin\Omega_{4-r}\}.
  \label{eq:half-space-support}
\end{align}
The subcomplex condition implies $\dd C_\Omega^r(\Lambda)\subset C_\Omega^{r+1}(\Lambda)$.
We introduce a dynamical two-form field $b\in C_\Omega^2(\Lambda)$ and restrict its one-form gauge transformations to
\begin{align*}
  a\longmapsto a+\lambda^1,\qquad
  b\longmapsto b+\dd\lambda^1,\qquad
  \lambda^1\in C_\Omega^1(\Lambda).
\end{align*}
These support conditions implement the Dirichlet boundary conditions of Ref.~\cite{Choi:2021kmx} for the new gauge field and its gauge transformations.

The half-space gauging sum is
\begin{align}
  Z_\Omega[w]
  =\Ncal_\Omega
  \sum_{a\in C^1(\Lambda)}\sum_{b\in C_\Omega^2(\Lambda)}
  \prod_{p\in\Lambda_2}w((\dd a)_p-b_p)
  \prod_{\substack{c\in\Lambda_3\\c^*\in\Omega_1}}
    \delta_{(\dd b)_c,0},
  \label{eq:half-space-gauging-sum}
\end{align}
where $\Ncal_\Omega$ is independent of the fields.
The fields are extended by zero outside their specified support when evaluating coboundaries.
In particular, $\dd b$ vanishes automatically for $c^*\notin\Omega_1$.

Split $a=a_{\mathrm{in}}+a_{\mathrm{out}}$, where $a_{\mathrm{in}}$ is supported on the links with $\ell^*\in\Omega_3$.
The change of variables $b'=b-\dd a_{\mathrm{in}}$ preserves the support condition and the flatness constraint on $b$ and gives
\begin{align*}
  \dd a-b=\dd a_{\mathrm{out}}-b'.
\end{align*}
Thus, the summand becomes independent of $a_{\mathrm{in}}$.
Equivalently, the allowed one-form gauge transformation sets precisely these original links to zero.
Below we write $a$ for $a_{\mathrm{out}}$ and retain all original links with $\ell^*\notin\Omega_3$, including those at the interface.

\subsection{Fourier transformation and the boundary phase}

Introduce an auxiliary dual link field $\at\in C^1(\Omega)$ to impose flatness:
\begin{align*}
  \delta_{(\dd b)_c,0}
  =\frac{1}{N}\sum_{\at_{c^*}\in\Zb_N}
    \chi_k(\at_{c^*},(\dd b)_c),
  \qquad c^*\in\Omega_1.
\end{align*}
This identity applies to every constraint, whether or not the constraints are independent.
Using Lemma~\ref{lemma:cochain-integration-by-parts} with $q=2$, we obtain
\begin{align*}
  \prod_{c^*\in\Omega_1}\chi_k(\at_{c^*},(\dd b)_c)
  =\prod_{p^*\in\Omega_2}\chi_k(-(\dd\at)_{p^*},b_p).
\end{align*}
For each $p^*\in\Omega_2$, put $f=(\dd a)_p$ and $h=(\dd\at)_{p^*}$.
The sum over $b_p$ is then
\begin{align}
  \sum_{b\in\Zb_N}w(f-b)\chi_k(-h,b)
  &=\sum_{u\in\Zb_N}w(u)\chi_k(-h,f-u)\notag\\
  &=\chi_k(-f,h)\sum_{u\in\Zb_N}w(u)\chi_k(h,u)\notag\\
  &=\sqrt{N}\,\wt(h)\chi_k(-f,h),
  \label{eq:half-space-plaquette-sum}
\end{align}
where $u=f-b$ and $\wt=\Fs_k w$.
This is the same shifted Fourier identity as Eq.~\eqref{eq:defect-forward-fourier}.
In the full-space calculation all original links could be eliminated, so that $f=0$; here the surviving boundary links produce the additional phase.

After summing over $b$, we can write the partition function as
\begin{align*}
  Z_\Omega[w]
  &=\widetilde{\Ncal}_\Omega
    \sum_{a_{\mathrm{out}}}\sum_{\at\in C^1(\Omega)}
    \left[\prod_{p^*\notin\Omega_2}w((\dd a)_p)\right]
    \left[\prod_{p^*\in\Omega_2}\wt((\dd\at)_{p^*})\right]
    \Phi_\Omega(a,\at),\\
  \Phi_\Omega(a,\at)
  &:=\prod_{p^*\in\Omega_2}
      \chi_k(-(\dd a)_p,(\dd\at)_{p^*}),
\end{align*}
where $\widetilde{\Ncal}_{\Omega}$ is a constant independent of the field configuration. The bulk weights are therefore $w$ outside and $\wt$ inside, as in Sec.~\ref{sec:duality-defects}.

\subsection{The tetrahedral-prism weight}

To determine the support of $\Phi_\Omega$, we expand the two coboundaries using the incidence coefficients of Appendix~\ref{app:cochain}.
\begin{align}
  \Phi_\Omega(a,\at)
  &=\prod_{\ell,c^*}\chi_k(a_\ell,\at_{c^*})^{m_{\ell c^*}(\Omega)},\notag\\
  m_{\ell c^*}(\Omega)
  &:=-\sum_{p^*\in\Omega_2}[p:\ell][p^*:c^*]
   =-\sum_{p^*\in\Omega_2}[\ell^*:p^*][p^*:c^*].
  \label{eq:half-space-boundary-product}
\end{align}
Here $\ell$ is an original link, $c^*$ is a dual link, and the last equality uses Eq.~\eqref{eq:dual-incidence-sign} with $q=1$.
If $\ell^*\in\Omega_3$, every face of $\ell^*$ belongs to $\Omega_2$, and the coefficient vanishes by $\partial^2\ell^*=0$.
If $c^*\notin\Omega_1$, it also vanishes by the subcomplex condition.
For each tetrahedral facet spanned by $\ell$ and $c^*$ in the covering lattice, $c^*$ is an edge of the dual cube $\ell^*$.
The two faces of this cube containing $c^*$ correspond to the two 16-cells sharing the facet.
Their incidence products have opposite signs in Eq.~\eqref{eq:half-space-boundary-product}.
For each interior facet of $R_\Omega$, the contributions from the two adjacent 16-cells cancel. Each facet separating a selected 16-cell from an unselected one contributes a factor with exponent $+1$ or $-1$.
In the periodic quotient these exponents are summed for each pair of link variables.

This cancellation also holds when the boundary is singular.
For example, in a sufficiently large periodic lattice, take $\Omega$ to consist of two coplanar dual plaquettes meeting only at a vertex, together with their faces.
Their 16-cells also meet only at that vertex, so $\partial R_\Omega$ consists of two three-spheres joined at one point.
When $\partial R_\Omega$ is an embedded closed piecewise-linear three-manifold, it serves as the interface $M$ in Sec.~\ref{sec:defect-geometry}.
Doubling each of its tetrahedral facets gives the prism of Fig.~\ref{fig:tetrahedralprism}.
The signs in Eq.~\eqref{eq:half-space-boundary-product} refer to the fixed reference orientations of the cochains.
Reading the two active link variables in the standard arrow directions of that figure gives the local weight
\begin{align}
  D(a,\at)=\chi_k(a,\at).
  \label{eq:half-space-prism-weight}
\end{align}

Eq.~\eqref{eq:half-space-prism-weight} is exactly the interface weight already given in Eq.~\eqref{dualitydefectweight}.
Thus, half-space gauging reproduces the same local coupling as the earlier 16-cell construction.
With an appropriate choice of the normalization $\Ncal_{\Omega}$, including the appropriate Euler counterterm \cite{Hayashi:2022fkw,Kaidi:2022cpf}, we can also reproduce the factors $s,l,\sba,\lb$ in Eq.~\eqref{dualitydefectweight}.

\section{Conclusion and discussion}
\label{sec:conclusion}

In this paper, we constructed topological duality interfaces in four-dimensional $\Zb_N$ lattice gauge theories with generalized plaquette actions.
For any even Fourier-dual pair $w$ and $\wt=\Fs_k w$, we specified local interface weights and proved their invariance under the allowed local deformations of the lattice.
At self-duality, these interfaces become defects representing non-invertible symmetries of a single theory.

The local construction gives identities that are independent of the detailed bulk dynamics.
In particular, we computed the expectation value of the duality interface on $S^3$, derived the connected-sum relation, and constructed center symmetry defects and their junctions with the duality interface.
We also derived crossing relations involving these center symmetry defects and junctions.
We further showed that half-space gauging with Dirichlet boundary conditions, following \cite{Choi:2021kmx}, reproduces the same interface.

An important direction is to use these non-invertible symmetries to study the phase structure.
Duality defects can impose restrictions on the infrared behavior.
In particular, non-invertible duality symmetry can obstruct a trivially gapped phase, and more general obstructions to symmetry-preserving gapped phases have been derived in Refs.~\cite{Choi:2021kmx,Apte:2022xtu}.
$\Zb_N$ lattice gauge theories have a vast parameter space, and it would be interesting to explore this space using non-invertible symmetries.

Such an analysis could complement numerical studies of generalized plaquette actions \cite{Fukugita:1982kk,Creutz:1982dn} and recent numerical studies of emergent photons in discrete gauge theories \cite{Giansiracusa:2025hfj}.
The self-dual parameter spaces found here provide concrete families in which to investigate whether the infrared theory is gapless or gapped, and whether the duality symmetry is preserved or spontaneously broken.
Combining the exact defect and junction relations with measurements of line operators and defect observables may help distinguish these possibilities and constrain the phase boundaries.
Determining which self-dual theories admit a continuum limit remains a dynamical question.

Another important direction is to extend the present defect construction to theories with dynamical matter.
Modified Villain formulations of four-dimensional $\U(1)$ lattice gauge theory admit both electric and magnetic matter and an exact duality exchanging the two sectors \cite{Sulejmanpasic:2019ytl}.
Extensions incorporating a $\theta$-term and dyonic matter have also been developed \cite{Anosova:2022cjm}.
In particular, Anosova, Gattringer, Iqbal, and Sulejmanpasic studied the phase structure of self-dual models with matched electric and magnetic bosonic matter, using a worldline representation and Monte Carlo simulations, and found evidence for spontaneous breaking of self-duality \cite{Anosova:2022yqx}.
These models realize invertible electric-magnetic duality in the presence of dynamical matter and provide useful guidance for choosing matter content and couplings compatible with self-duality.

For the $\Zb_N$ theories considered here, charged matter reduces the electric one-form symmetry to the subgroup of the center that acts trivially on all matter fields \cite{Gaiotto:2014kfa}.
For a single matter charge $q$, this subgroup is $\Zb_{\gcd(N,q)}$; in particular, fundamental matter breaks the electric center symmetry completely.
Thus, extending our construction by half-space gauging requires identifying a surviving one-form symmetry to gauge and determining how the matter couplings transform under duality.
This question is also relevant to the relation between confinement and Higgs regimes in lattice gauge theories \cite{Osterwalder:1977pc,Fradkin:1978dv}.
A useful related example is the Cardy--Rabinovici model \cite{Cardy:1981qy,Cardy:1981fd}, a four-dimensional $\U(1)$ gauge theory with electric and magnetic matter.
Non-invertible self-duality defects in this model have been constructed in Ref.~\cite{Hayashi:2022fkw}.
Building on these examples, we could identify matter couplings compatible with self-duality in $\Zb_N$ lattice gauge theories and extend our local construction of non-invertible duality defects and their junctions to these theories.

An important direction for future work is to develop a Hamiltonian description of the non-invertible duality symmetries constructed here.
Previous Hamiltonian studies of $(3+1)$-dimensional lattice gauge theories with discrete gauge groups include
Refs.~\cite{Horn:1979fy,Harlow:2018tng,Choi:2021kmx,Delcamp:2021szr,Mariani:2023eix,Moradi:2023dan,Gorantla:2024ocs,Gorantla:2024ptu,Hofmeier:2026pwm}.

\subsection*{Acknowledgments}
The author would like to thank Hidenori Fukaya, Naoto Kan, Kimyeong Lee, Tetsuya Onogi, Qiaoyin Pan, and Hossein Yavartanoo for useful discussions and comments.
The author completed this work while visiting the Yau Mathematical Sciences Center at Tsinghua University and the Beijing Institute of Mathematical Sciences and Applications.
He is grateful to both institutions for their hospitality.
This work was supported in part by JSPS KAKENHI Grant Number JP26K07086.

\appendix

\section{Summary of cochain notation}
\label{app:cochain}

Here, we summarize the cochain notation used to describe $\Zb_N$ lattice gauge theory.
All cochains have coefficients in $\Zb_N$, which we omit from the notation, and all sums are understood modulo $N$.

Let $\Lambda$ be a finite four-dimensional hypercubic lattice with periodic boundary conditions.
For $q=0,\ldots,4$, let $\Lambda_q$ denote the set of $q$-cells with chosen reference orientations.
A $q$-cochain $c$ assigns a value $c_\sigma\in\Zb_N$ to each $\sigma\in\Lambda_q$, extended to the opposite orientation by $c_{-\sigma}=-c_\sigma$.
The cochains form an abelian group under pointwise addition, denoted by
\begin{align}
  C^q(\Lambda):=\{c:\Lambda_q\to\Zb_N\}.
  \label{eq:cochain-group}
\end{align}
In particular, $0$-, $1$-, and $2$-cochains assign variables to sites, links, and plaquettes, respectively.

For $q=0,\ldots,3$, the coboundary homomorphism $\dd:C^q(\Lambda)\to C^{q+1}(\Lambda)$ is defined by
\begin{align}
  (\dd c)_\sigma:=\sum_{\tau\in\partial\sigma}c_\tau,
  \qquad \sigma\in\Lambda_{q+1},
  \label{eq:coboundary}
\end{align}
where the boundary faces are counted with their induced orientations and multiplicities.
For example, for $\lambda\in C^0(\Lambda)$ and $a\in C^1(\Lambda)$,
\begin{align}
  (\dd\lambda)_{\expval{ij}}=\lambda_j-\lambda_i,
  \qquad
  (\dd a)_p=\sum_{\ell\in\partial p}a_\ell=f_p.
  \label{eq:coboundary-examples}
\end{align}
Thus, the gauge transformation~\eqref{eq:gauge-transformation} is written as $a\mapsto a+\dd\lambda$.
Since the oriented boundary of a boundary vanishes, $\dd^2=0$, and these groups form the cochain complex
\begin{align}
  0\longrightarrow C^0(\Lambda)
  \xrightarrow{\dd}C^1(\Lambda)
  \xrightarrow{\dd}\cdots
  \xrightarrow{\dd}C^4(\Lambda)
  \longrightarrow0.
  \label{eq:cochain-complex}
\end{align}

Let $\Lambda^*$ be the dual lattice of $\Lambda$.
There is a canonical bijection between the underlying $r$-cells of $\Lambda$ and $(4-r)$-cells of $\Lambda^*$, denoted by $\sigma\mapsto\sigma^*$.
Fix the standard orientation of the four-torus, and orient each dual cell so that an oriented tangent basis of $\sigma$ followed by one of $\sigma^*$ agrees with the ambient orientation.
Thus, the reference orientations on the dual lattice are determined by those on $\Lambda$.
For the bicharacter $\chi_k$ defined in Eq.~\eqref{bichar}, we have the following discrete integration-by-parts identity.
\begin{lemma}[Discrete integration by parts]\label{lemma:cochain-integration-by-parts}
  For $q=0,\ldots,3$, $b\in C^q(\Lambda)$, and $g\in C^{3-q}(\Lambda^*)$, we have
  \begin{align}
    \prod_{\sigma\in\Lambda_{q+1}}
      \chi_k\bigl((\dd b)_\sigma,g_{\sigma^*}\bigr)
    =\prod_{\tau\in\Lambda_q}
      \chi_k\bigl(b_\tau,(-1)^{q+1}(\dd g)_{\tau^*}\bigr).
    \label{eq:cochain-integration-by-parts}
  \end{align}
\end{lemma}
\begin{proof}
Let $[\sigma:\tau]$ denote the coefficient of the oriented $q$-cell $\tau$ in $\partial\sigma$.
The dual orientation convention gives
\begin{align}
  [\tau^*:\sigma^*]=(-1)^{q+1}[\sigma:\tau].
  \label{eq:dual-incidence-sign}
\end{align}
Indeed, the outward normal of $\sigma$ at $\tau$ and that of $\tau^*$ at $\sigma^*$ point in opposite directions.
This contributes a minus sign, while moving the normal past the $q$ tangent directions of $\tau$ contributes $(-1)^q$.
The relation holds for each local incidence, and hence also after summing the multiplicities arising from periodic identifications.

Expanding the coboundary and using the bicharacter property, we obtain
\begin{align*}
  \prod_{\sigma\in\Lambda_{q+1}}
    \chi_k\bigl((\dd b)_\sigma,g_{\sigma^*}\bigr)
  &=\prod_{\sigma\in\Lambda_{q+1}}\prod_{\tau\in\Lambda_q}
    \chi_k(b_\tau,g_{\sigma^*})^{[\sigma:\tau]}\\
  &=\prod_{\tau\in\Lambda_q}
    \chi_k\left(b_\tau,
      \sum_{\sigma\in\Lambda_{q+1}}[\sigma:\tau]g_{\sigma^*}\right)\\
  &=\prod_{\tau\in\Lambda_q}
    \chi_k\bigl(b_\tau,(-1)^{q+1}(\dd g)_{\tau^*}\bigr),
\end{align*}
where the last equality uses Eq.~\eqref{eq:dual-incidence-sign} and the definition of the dual coboundary.
Since the lattice is periodic, there is no boundary term.
\end{proof}
The sign in Eq.~\eqref{eq:cochain-integration-by-parts} is positive for $q=1,3$ and negative for $q=0,2$.
For $N=2$, the sign is immaterial.

\section{Evaluation of the Gauss sum}
\label{app:GaussSum}
We can determine $d_{N,k}$ for every pair of coprime integers $N>1$ and
$k$. To do so, we evaluate the Gauss sum:
\begin{align}
  G(k;N):=\sum_{r=0}^{N-1}e^{\frac{2\pi i k r^2}{N}}.
\end{align}
Since $G(k;N)$ depends only on $k$ modulo $N$, we take
$1\leq k<N$.
For an odd positive integer $m$, let
$\bigl(\frac{a}{m}\bigr)$ denote the Jacobi symbol, with the convention
$\bigl(\frac{a}{1}\bigr)=1$.  We use the properties of Jacobi symbols in
Ref.~\cite{Apostol}, Sec.~9.7 (Theorems 9.9--9.11), and the quadratic
Gauss-sum evaluations and reciprocity relation in Sec.~9.10 of the same
reference (Eqs.~(29), (30) and Theorem 9.16).

\subsection{Odd \texorpdfstring{$N$}{N}}
First, consider the case in which $N$ is odd. We begin with $N=p$, where $p$ is an odd prime and $p\nmid k$.
Eq.~(29) of Ref.~\cite{Apostol} states that
\begin{align}
  G(k;p)=\left(\frac{k}{p}\right)G(1;p).
  \label{eq:Gauss-prime-character}
\end{align}

Next let us consider $N=p^{\alpha}$ for $\alpha \geq 2$ and an odd prime $p$.  We can write each residue modulo $p^\alpha$ uniquely as
$r=s+t p^{\alpha-1}$, where $0\leq s<p^{\alpha-1}$ and $0\leq t<p$.
The term $kt^2p^{\alpha-2}$ in $kr^2/p^\alpha$ is an integer, so
\begin{align}
  G(k;p^\alpha)
  &=\sum_{s=0}^{p^{\alpha-1}-1}
    \sum_{t=0}^{p-1}e^{2\pi i k (s+t p^{\alpha-1})^2/p^\alpha}
  =\sum_{s=0}^{p^{\alpha-1}-1}e^{2\pi i ks^2/p^\alpha}
    \sum_{t=0}^{p-1}e^{2\pi i(2ks)t/p},\\
  \sum_{t=0}^{p-1}e^{2\pi i(2ks)t/p}
  &=\begin{cases}
    p,&p\mid s,\\
    0,&p\nmid s.
  \end{cases}
\end{align}
The identity in the second line follows from the finite geometric series and
$\gcd(2k,p)=1$. Only $s=pu$, with $0\leq u<p^{\alpha-2}$,
therefore contributes, giving
\begin{align}
  G(k;p^\alpha)
  =p\sum_{u=0}^{p^{\alpha-2}-1}e^{2\pi i ku^2/p^{\alpha-2}}.
\end{align}
This proves the recurrence in Ref.~\cite[Chapter~8, Exercise~16(b)]{Apostol}:
\begin{align}
  G(k;p^\alpha)=p\,G(k;p^{\alpha-2}),
  \qquad G(k;1):=1,
  \label{eq:Gauss-prime-power-recurrence}
\end{align}
where the convention for modulus one also covers $\alpha=2$.
Iterating this recurrence yields
\begin{align}
  G(k;p^\alpha)
  =\begin{cases}
    p^{\alpha/2},&\alpha\ \text{even},\\
    p^{(\alpha-1)/2}G(k;p),&\alpha\ \text{odd}.
  \end{cases}
\end{align}
Applying the same formula with $k=1$ and using
Eq.~\eqref{eq:Gauss-prime-character}, together with
$\bigl(\frac{k}{p}\bigr)=\pm1$, we obtain
\begin{align}
  G(k;p^\alpha)
  =\left(\frac{k}{p}\right)^\alpha G(1;p^\alpha).
  \label{eq:Gauss-prime-power-character}
\end{align}

To combine distinct prime powers, let $m,n$ be coprime positive
integers. The Chinese remainder theorem implies that
$r=nx+my$, with $0\leq x<m$ and $0\leq y<n$, runs through every
residue modulo $mn$ exactly once: multiplication by $n$ modulo $m$
and by $m$ modulo $n$ are both invertible. Since
\begin{align}
  \frac{k(nx+my)^2}{mn}
  =\frac{knx^2}{m}+2kxy+\frac{kmy^2}{n},
\end{align}
the integer cross term $2kxy$ contributes a phase of one. Hence
\begin{align}
  G(k;mn)
  &=\sum_{x=0}^{m-1}\sum_{y=0}^{n-1}
    e^{2\pi i knx^2/m}e^{2\pi i kmy^2/n}\\
  &=\left(\sum_{x=0}^{m-1}e^{2\pi i knx^2/m}\right)
    \left(\sum_{y=0}^{n-1}e^{2\pi i kmy^2/n}\right).
\end{align}
We have thus derived the factorization formula stated in
Ref.~\cite[Chapter~8, Exercise~16(a)]{Apostol},
\begin{align}
  G(k;mn)=G(kn;m)\,G(km;n).
  \label{eq:Gauss-CRT-factorization}
\end{align}
This identity holds for every integer $k$; only $\gcd(m,n)=1$ is
needed.

Finally, write the odd modulus as $N=\prod_j q_j$, where
$q_j=p_j^{\alpha_j}$ are powers of distinct odd primes, and set
$M_j=N/q_j$. Iterating Eq.~\eqref{eq:Gauss-CRT-factorization} gives
\begin{align}
  G(k;N)=\prod_j G(kM_j;q_j).
\end{align}
Because $p_j\nmid kM_j$, Eq.~\eqref{eq:Gauss-prime-power-character}
and multiplicativity of the Legendre symbol imply
\begin{align}
  G(kM_j;q_j)
  &=\left(\frac{kM_j}{p_j}\right)^{\alpha_j}G(1;q_j)\\
  &=\left(\frac{k}{p_j}\right)^{\alpha_j}
    \left(\frac{M_j}{p_j}\right)^{\alpha_j}G(1;q_j)
   =\left(\frac{k}{p_j}\right)^{\alpha_j}G(M_j;q_j).
\end{align}
The product of $\bigl(\frac{k}{p_j}\bigr)^{\alpha_j}$ is
$\bigl(\frac{k}{N}\bigr)$ by the definition of the Jacobi symbol,
while $\prod_j G(M_j;q_j)=G(1;N)$ by the same factorization with
$k=1$. Consequently, for every odd $N$ with $\gcd(k,N)=1$,
\begin{align}
  G(k;N)=\left(\frac{k}{N}\right)G(1;N).
  \label{eq:Gauss-odd-general}
\end{align}
Using the classical evaluation of $G(1;N)$ in Eq.~(30) of
Ref.~\cite{Apostol}, we obtain
\begin{align}
  \frac{G(k;N)}{\sqrt N}
  =\varepsilon_N\left(\frac{k}{N}\right),
  \qquad
  \varepsilon_m:=
  \begin{cases}
    1,&m\equiv1\pmod4,\\
    i,&m\equiv3\pmod4,
  \end{cases}
  \quad (N\ \text{odd}).
  \label{eq:Gauss-odd-Jacobi}
\end{align}

\subsection{Even \texorpdfstring{$N$}{N}}
Next let $N\equiv2\pmod4$. Writing $N=2m$ with $m$ odd,
we note that $k$ is necessarily odd. The terms in the Gauss sum
with $r$ and $r+m$ cancel, since
\begin{align}
  \exp\!\left(\frac{2\pi i k(r+m)^2}{2m}\right)
  =-\exp\!\left(\frac{2\pi i kr^2}{2m}\right).
\end{align}
Consequently, $G(k;N)=0$ in this case.

Finally, consider $4\mid N$, so that $k$ is again odd. We use the
quadratic reciprocity formula of Ref.~\cite{Apostol},
Sec.~9.10, Theorem 9.16. In its notation, define
\begin{align}
  S(a,m):=\sum_{r=0}^{m-1}\exp\!\left(\frac{\pi i a r^2}{m}\right).
\end{align}
For positive integers \(a\) and \(m\) with \(am\) even, the theorem gives
\begin{align}
  S(a,m)=e^{\pi i/4}\sqrt{\frac{m}{a}}\,
  S(m,a)^{*}.
\end{align}
For odd $k$, the residues $r=2s$ and $r=2s+k$ modulo $2k$,
with $0\leq s<k$, cover the even and odd classes, respectively.
Using the $2k$-periodicity of the summand, we therefore obtain
\begin{align}
  S(N,2k)
  &=\sum_{s=0}^{k-1}\left[
    e^{2\pi i Ns^2/k}+e^{\pi i N(2s+k)^2/(2k)}\right]
    \notag\\
  &=\left(1+e^{\pi i Nk/2}\right)G(N;k)
   =2G(N;k)\qquad(4\mid N).
\end{align}
Here the mixed term in $(2s+k)^2$ contributes $e^{2\pi i Ns}=1$,
and $e^{\pi i Nk/2}=1$ since $4\mid N$.
Combining this with $S(2k,N)=G(k;N)$ and the reciprocity formula,
we find
\begin{align}
  G(k;N)
  &=(1+i)\sqrt{\frac Nk}\,G(N;k)^{*}
    \notag\\
  &=(1+i)\sqrt N\,\varepsilon_k^{-1}
    \left(\frac{N}{k}\right).
  \label{eq:Gauss-even-Jacobi}
\end{align}
In the second line, Eq.~\eqref{eq:Gauss-odd-Jacobi} was applied to
the odd modulus $k$, including the convention $k=1$.
In particular, although the imaginary part depends on $k\pmod4$,
$\operatorname{Re}[(1+i)\varepsilon_k^{-1}]=1$ for every odd $k$.

\subsection{Summary of the result}
To summarize, the real part relevant for the dimension formula is
\begin{align}
  \frac{1}{\sqrt N}\re G(k;N)
  =\begin{cases}
    \displaystyle\left(\frac{k}{N}\right),
       &N\equiv1\pmod4,\\[4pt]
    \displaystyle\left(\frac{N}{k}\right),
       &N\equiv0\pmod4,\\[4pt]
    0, &N\equiv2,3\pmod4.
  \end{cases}
  \label{eq:Gauss-real-general}
\end{align}
Substitution into the expression for $d_{N,k}$ gives the exact result
\begin{align}
  d_{N,k}=\floor{\frac N4}-
  \begin{cases}
    \displaystyle\frac{1}{2}\left[1-\left(\frac{k}{N}\right)\right],
      &N\equiv1\pmod4,\\[6pt]
    \displaystyle\frac{1}{2}\left[1-\left(\frac{N}{k}\right)\right],
      &N\equiv0\pmod4,\\[6pt]
    0, &N\equiv2,3\pmod4.
  \end{cases}
  \label{eq:dNk-general-Jacobi}
\end{align}
In particular, $d_{N,1}=\floor{N/4}$ for all $N>1$.
The dependence on $k$ disappears for $N\equiv2,3\pmod4$ and
also whenever $N$ is a perfect square. For $N\equiv0,1\pmod4$
with $N$ not a perfect square, the Jacobi factor defines a nontrivial
quadratic character on $(\Zb/N\Zb)^\times$. Therefore, precisely
half of the $\varphi(N)$ admissible values of $k$ yield
$d_{N,k}=\floor{N/4}$, while the other half yield
$d_{N,k}=\floor{N/4}-1$.

The Jacobi symbols in Eq.~\eqref{eq:dNk-general-Jacobi}
can be evaluated using the following standard identities
\cite[Sec.~9.7]{Apostol}.
Here $m$ is odd and positive; the last identity additionally
assumes that $a$ is odd, positive, and coprime to $m$.
\begin{align}
  \left(\frac{-1}{m}\right)&=(-1)^{(m-1)/2},\qquad
  \left(\frac{2}{m}\right)=(-1)^{(m^2-1)/8},
  \notag\\
  \left(\frac{a}{m}\right)
  &=(-1)^{(a-1)(m-1)/4}\left(\frac{m}{a}\right).
\end{align}
The Jacobi symbol is multiplicative in both arguments and
depends only on the residue class of its numerator modulo
its denominator. Thus, for odd positive integers $m,n$
and arbitrary integers $a,b$, we have
\begin{align}
  \left(\frac{ab}{m}\right)
  =\left(\frac{a}{m}\right)
  \left(\frac{b}{m}\right),\qquad
  \left(\frac{a}{mn}\right)=
  \left(\frac{a}{m}\right)
  \left(\frac{a}{n}\right),\qquad
  \left(\frac{a}{m}\right)=
  \left(\frac{a+bm}{m}\right).
\end{align}
Together with the reciprocity and supplementary laws above,
these identities allow Jacobi symbols to be evaluated
efficiently without prime factorization.
For example, $d_{5,2}=0$ while $d_{5,1}=1$, and
$d_{8,3}=1$ while $d_{8,1}=2$.

For completeness, Table~\ref{tab:dNk-exceptions-30} lists every
exceptional value with $d_{N,k}=\floor{N/4}-1$ for $2\le N\le30$.
For all remaining coprime pairs in this range,
$d_{N,k}=\floor{N/4}$.
\begin{table}[htbp]
\centering
\small
\begin{tabular}{c l}
\hline
$N$ & $k$ such that $d_{N,k}=\floor{N/4}-1$ \\
\hline
$5$ & $2,\ 3$ \\
$8$ & $3,\ 5$ \\
$12$ & $5,\ 7$ \\
$13$ & $2,\ 5,\ 6,\ 7,\ 8,\ 11$ \\
$17$ & $3,\ 5,\ 6,\ 7,\ 10,\ 11,\ 12,\ 14$ \\
$20$ & $3,\ 7,\ 13,\ 17$ \\
$21$ & $2,\ 8,\ 10,\ 11,\ 13,\ 19$ \\
$24$ & $7,\ 11,\ 13,\ 17$ \\
$28$ & $5,\ 11,\ 13,\ 15,\ 17,\ 23$ \\
$29$ & $2,\ 3,\ 8,\ 10,\ 11,\ 12,\ 14,\ 15,\ 17,\ 18,\ 19,\ 21,\ 26,\ 27$ \\
\hline
\end{tabular}
\caption{All exceptional pairs with $2\le N\le30$ and
  $1\le k<N$, $\gcd(N,k)=1$. For any admissible pair not listed,
  $d_{N,k}=\floor{N/4}$.}
\label{tab:dNk-exceptions-30}
\end{table}

\bibliographystyle{utphys}
\bibliography{ref}
\end{document}